\documentclass[twoside,11pt]{article}

\usepackage{jmlr2e}

\usepackage{booktabs}
\usepackage{subfig}
\usepackage{float}
\usepackage{amssymb}
\usepackage{graphicx}
\usepackage{fancyref}
\usepackage{longtable}
\usepackage{amsmath}
\usepackage{longtable}
\usepackage{amsfonts}
\usepackage{bm}
\usepackage{lmodern}
\usepackage{enumitem}
\ShortHeadings{gGOF for Correlated Data}{Zhang and Wu}
\firstpageno{1}

\begin{document}

\title{Generalized Goodness-Of-Fit Tests for Correlated Data}

\author{\name Hong Zhang \email hzhang@wpi.edu \\
       \addr Department of Mathematical Sciences\\
       Worcester Polytechnic Institute\\
       Worcester, MA 01609, USA
       \AND
       \name Zheyang\ Wu \email zheyangwu@wpi.edu \\
       \addr Department of Mathematical Sciences\\
       Worcester Polytechnic Institute\\
       Worcester, MA 01609, USA}

\editor{}

\maketitle

\begin{abstract}
This paper concerns the problem of applying the generalized goodness-of-fit (gGOF) type tests for analyzing correlated data. The gGOF family broadly covers the maximum-based testing procedures by ordered input $p$-values, such as the false discovery rate procedure, the Kolmogorov-Smirnov type statistics, the $\phi$-divergence family, etc. Data analysis framework and a novel $p$-value calculation approach is developed under the Gaussian mean model and the generalized linear model (GLM). We reveal the influence of data transformations to the signal-to-noise ratio and the statistical power under both sparse and dense signal patterns and various correlation structures. In particular, the innovated transformation (IT), which is shown equivalent to the marginal model-fitting under the GLM, is often preferred for detecting sparse signals in correlated data. We propose a testing strategy called the digGOF, which combines a \underline{d}ouble-adaptation procedure (i.e., adapting to both the statistic's formula and the truncation scheme of the input $p$-values) and the \underline{I}T within the \underline{gGOF} family. It features efficient computation and robust adaptation to the family-retained advantages for given data. Relevant approaches are assessed by extensive simulations and by genetic studies of Crohn's disease and amyotrophic lateral sclerosis. Computations have been included into the R package {\it SetTest} available on CRAN.
\end{abstract}


\section{Introduction}

With a long history and numerous applications, the goodness-of-fit (GOF) tests are one of the breakthroughs of statistical data analysis algorithms \citep{komogorov1933, kotz2012breakthroughs}. In particular, the GOF tests provide a promising tool for signal detection problems in analyzing big data. A collection of the GOF statistics such as the Higher Criticism (HC) type statistics, the Berk-Johns (BJ) type statistics, and the $\phi$-divergence statistics, have been proven asymptotically optimal for weak-and-rare signals  \citep{Donoho2004, jager2007goodness, Wu2014detection}. For a unified study of these GOF statistics, we proposed a general family of statistics called the gGOF, which is defined by a generic functional and a general truncation scheme of input $p$-values $P_1, ..., P_n$ \citep{Zhang2016distributions}. Specifically, under the null hypothesis, all continuous $p$-values are from Uniform$(0, 1)$. Let $P_{(1)} \leq ... \leq P_{(n)}$ be the ordered $p$-values. A  gGOF statistic measures the supremum departure of $P_{(i)}$ from its null expectation, which is roughly $\frac{i}{n}$:
\begin{equation}
\label{equ.gGOFstat}
\begin{array}{ll}
	S_{n, f, \mathcal{R}} = \sup_{\mathcal{R}} f(\frac{i}{n}, P_{(i)}), 
\end{array}
\end{equation}
where at a fixed $x=\frac{i}{n}$ the function $f(x, y)$ is monotonically decreasing in $y=P_{(i)}$, and $\mathcal{R}$ represents an arbitrary truncation scheme for these input $p$-values.  The only requirement for the gGOF is the monotonicity of the function $f$ so that the smaller the input $p$-values the larger the statistic, and the stronger the evidence is to against the null hypothesis. Since a smaller $p$-value should always be more likely to represent a signal, this monotonicity is essentially the minimal requirement for the efficiency of such maximum-based statistics of ordered $p$-values.  The truncation scheme $\mathcal{R}$ can be based on the index $i$ and/or the magnitude of $P_{(i)}$, for example, $\mathcal{R} = \{i:  k_0 \leq i \leq k_1 \} \bigcap \{P_{(i)}:  \alpha_0 \leq P_{(i)} \leq \alpha_1 \}$ for given $k_0 \leq k_1 \in \{1, ..., n\}$ and $\alpha_0 \leq \alpha_1 \in [0, 1]$. $\mathcal{R}$ can be even more general, for example, by taking sets of non-adjacent ordered $p$-values. 

Moreover, the monotonicity of $f$ is equivalent to the fact that $f$ is invertible and thus the distribution of a gGOF statistic can be written as a probability function of the cross-boundaries of the ordered $p$-values:
\begin{equation}
\label{equ.cross.bound.prob}
\mathbb{P}(S_{n, f, \mathcal{R}} \leq b) 
=\mathbb{P}(\sup_{\mathcal{R}} f(\frac{i}{n}, P_{(i)}) \leq b)  
=\mathbb{P}(P_{(i)}>u_i(b) \text{, for all } i \text{ and } P_{(i)} \in \mathcal{R}),
\end{equation}
where the boundaries are decided by the inverse of $f$ and the threshold $b$:
\begin{equation}
\label{equ.rejectionBoundary}
u_i(b) = f^{-1}(\frac{i}{n}, b).
\end{equation}
 
Under the assumption that the input $p$-values are independent and identically distributed ({\it iid}), the calculations for the gGOF's $p$-values and statistical power have been well resolved \citep{Zhang2016distributions}. However, correlation is a ubiquitous phenomenon in real data analysis. A few problems need to be addressed for analyzing correlated data. First, to practically apply the gGOF statistics, analytical $p$-value calculation under arbitrary correlation is desired. Analytical calculation has advantages over the simulation-based empirical approaches from both methodological and computational perspectives. Secondly, to study the power of the gGOF for correlated data, it is important to understand how signal patterns and correlation structures would influence the signal-to-noise ratio (SNR). Proper transformations for the correlated $p$-values have a potential to advance signal detection, whereas improper ones could cause harm too. Thirdly, since different test statistics in the gGOF could have relative advantages under various signal patterns and data properties, it is ideal to fully utilize the family-retained advantages to provide a powerful and robust strategy. 

\begin{figure}[!t] 
	\begin{center}
		\subfloat{\includegraphics[width=2.5in,height=2.5in]{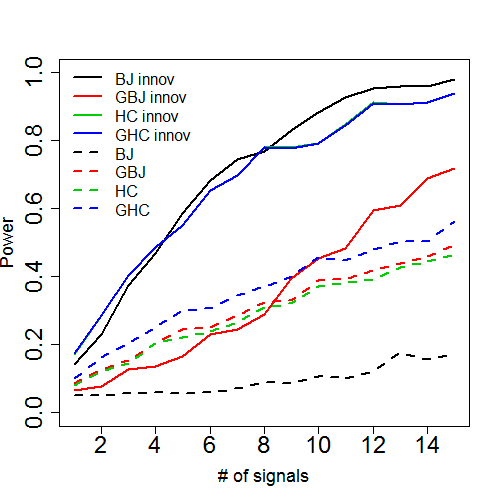}}
		\subfloat{\includegraphics[width=2.5in,height=2.5in]{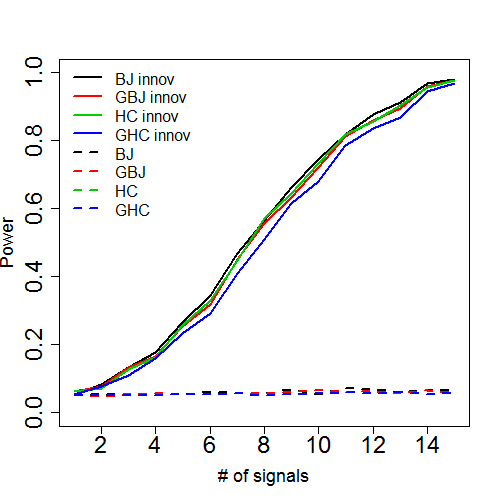}}\\
		\subfloat{\includegraphics[width=2.5in,height=2.5in]{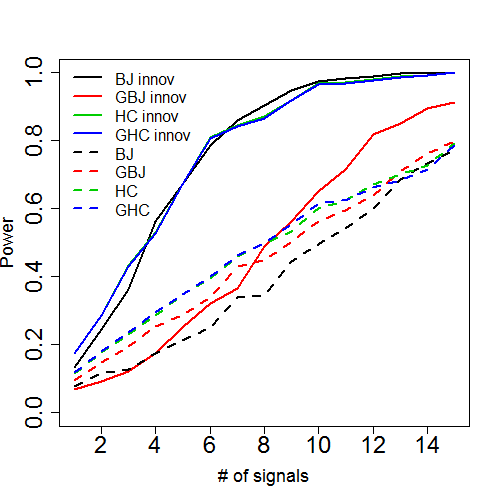}}
		\subfloat{\includegraphics[width=2.5in,height=2.5in]{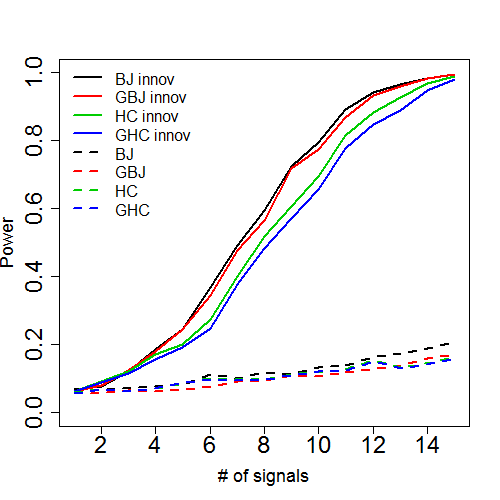}}
	\end{center}
	\caption{Statistical power for the HC, the GHC, the BJ, and the GBJ for detecting multiple signals with and without the innovated transformation (innov). Input test statistics $(T_1, ..., T_{100}) \sim N(\mu, \Sigma)$, where $\mu_j=A$ for $j\in M^\star = \{j_1, ..., j_K \}$ and $\mu_j = 0$ otherwise.  X-axis: the number of signals $K$. Top-left penal: $\Sigma^{(1)}_{ij}=0.3$ for $i\neq j$, $A=2$ (the curves of the HC-innov and the GHC-innov almost overlapped); top-right: $\Sigma^{(2)}_{ij}=-0.01$  for $i\neq j$, $A=0.4$; bottom-left: $\Sigma^{(3)}_{ij}=|i-j|^{-1}$  for $i\neq j$, $A=2$; bottom-right: $\Sigma^{(4)}= (\Sigma^{(3)})^{-1}$, $A=1$.  Type I error rate 0.05; 10,000 simulations; $M^\star$ is uniformly distributed on $\{1, ..., n\}$.}
\label{fig:power_GM_sig_number}
\end{figure}

In recent years, individual modifications have been proposed based on the gGOF statistics for analyzing correlated data. In particular, the GHC and the GBJ \citep{barnett2016generalized, sun2017set} were proposed based on the original HC and BJ statistics in the context of genetic association studies. These developments are very interesting and do improve statistical power under circumstances. However, the original HC and BJ could still have higher statistical power in analyses of correlated data. Figure \ref{fig:power_GM_sig_number} gives a few examples, for which the HC/BJ could tie or even surpass the GHC/GBJ in power, especially after the innovated transformation (IT). Detailed study and more comparisons of their relative advantages are given in Section \ref{Sect.CompPower}.

For analyzing correlated data, instead of carrying out individual developments of specific statistics, the idea of this paper is to address the whole statistic family, and allow data to automatically choose the best statistic function $f$ and truncation domain $\mathcal{R}$. The gGOF family is very broad, it covers not only the HC/BJ/GHC/GBJ type statistics, but also other testing procedures such as the Bonferroni procedure and the false discovery rate (FDR) procedure. Different statistics possess high statistical power under different signal patterns. For example, the initial HC statistic \citep{Donoho2004} is more powerful for sparser signals, while the reverse HC statistic \citep{Donoho2008} are more powerful for denser signals  \citep{Zhang2016distributions}. Moreover, the $p$-value truncation scheme $\mathcal{R}$ is also quite relevant. For example, the modified HC with $\mathcal{R} = \{ 1 < i \leq n/2, P_{(i)} \geq 1/n \}$ improves the performance over the statistic that has no truncations applied \citep{Donoho2004, Li2014higher}.  Thus, by allowing a general $f$ and $\mathcal{R}$ the gGOF family could retain all of these advantages to gain high and robust power in analyzing various data. 

In addition, the nature of the gGOF in (\ref{equ.cross.bound.prob})  leads to a unified calculation of $p$-values both for any given statistics and for the data-adaptive omnibus test. It is convenient to analytically calculate the exact null distribution, rather than relying on simulations, which are typically needed for omnibus tests \citep{barnett2016generalized, sun2017set}. Here we consider a double-adaptation test adapting to both $f$ and $\mathcal{R}$. Specifically, given a  gGOF statistic $S_{f, \mathcal{R}}$ at a fixed $n$, let $G_{f, \mathcal{R}}$ denote its survival function under the null hypothesis. The statistic of the double-adaptation test is the smallest $p$-value among all statistics indexed by both $f$ and $\mathcal{R}$, which represents the strongest statistical evidence against the null:
\begin{equation}
\label{equ.doubleomnibus}
S_o =  \inf_{f, \mathcal{R}} G_{f, \mathcal{R}}(S_{f, \mathcal{R}}). 
\end{equation}
The survival function of $S_o$ is 
\begin{equation}
\label{equ.cross.bound.prob.omni}
\mathbb{P}(S_o > s_o) =\mathbb{P}(S_{f, \mathcal{R}} \leq G_{f, \mathcal{R}}^{-1}(s_o), \text{ for all } f, \mathcal{R}) = \mathbb{P}(P_{(1)}>u^\star_1(s_o),...,P_{(n)}>u^\star_n(s_o)),
\end{equation}
where for each $i=1,...,n$,  
\begin{equation}
u^\star_i(s_o) = \sup_{f, \mathcal{R}} u_{f, \mathcal{R}, i} =  \sup_{f, \mathcal{R}} f^{-1}(\frac{i}{n}, G_{f, \mathcal{R}}^{-1}(s_o)). 
\end{equation}

The calculations for (\ref{equ.cross.bound.prob}) and (\ref{equ.cross.bound.prob.omni}) are unified under the general problem of calculating the cross-boundary probability. For independent data, such problem has been extensively studied \citep{Zhang2016distributions}.  In paper we will provide analytical calculation methods under correlated data. This methodology development has several advantages over the moment-matching (MM) strategy based on the extended beta-binomial distribution \citep{barnett2016generalized, sun2017set}. First, we provide direct derivation of the probability rather than the approximation by matching the moments of a surrogate distribution. Such explicit strategy allows obtaining the exact calculation under given correlation structures (e.g., the equal-correlation). Since the probability is an explicit function of correlations, this methods is more helpful for understanding how correlations influence the testing performance.  Secondly, the new calculation allows arbitrary truncation $\mathcal{R}$ for $p$-values. Thirdly, the computation is simpler. For example, under $\mathcal{R} = \{i:  k_0 \leq i \leq k_1 \}$, based on \citep{Zhang2016distributions}  our computational complexity is $O((k_1-k_0)^2)$ whereas the MM always needs $O(n^3)$. Moreover, the explicit calculation strategy shows better accuracy under broad circumstances.  

One complication in analyzing correlated data is that not only signal patterns but also correlation structures influence testing performance. Sometimes correlation information could be properly utilized for improving statistical power. To address this issue under the gGOF, we derive how the signal-to-noise ratios (SNRs, especially the largest ones) are changed through linear transformation of the input statistics based on the correlation matrix. In particular, we study the de-correlation transformation (DT) and the innovated transformation (IT) \citep{Hall2010, jin2014rare, fan2013optimal}, and reveal conditions for them to strengthen or weaken the SNR under the GMM and the GLM. For the GLM, we show that the marginal model-fitting is essentially the IT of the joint model-fitting. This result is interesting because it indicates that the computationally-simple marginal fitting is actually often superior to the computationally-expensive joint fitting for correlated data analysis. 
Because of this advantage, the IT could be applied before the gGOF, either implicitly through model-fitting or explicitly by transformation. As a natural extension of the iHC \citep{Hall2010}, we call such testing procedure the igGOF test. When the  double-adaptation omnibus test  in (\ref{equ.doubleomnibus}) is also carried out, we call the testing procedure the digGOF.

The remainder of the paper is organized as follows. Section \ref{Sect.Formulation} formulates the hypothesis testing problem under the GMM and the GLM. Section \ref{Sect.gGOF} presents some important testing statistics and procedures as examples of the gGOF. The analytical $p$-value calculation approaches are derived. A rejection boundary analysis is carried out to compare the performance of typical gGOF tests. Section \ref{Sect.transformations} presents a theoretical study on how the DT and the IT change SNR under the GMM and the GLM.  For weak-and-rare signals, the signal detection boundary and the asymptotic optimality of the iHC are given under the GLM. For more general signal and correlation patterns, we also provide a comprehensive study under the bivariate regression model. Section \ref{Sect.numerics} presents numerical  studies to evidence the accuracy of our $p$-value calculation approaches, and to compare the statistical power of relevant tests in the context of data transformations. Section \ref{Sect.GWAS} illustrates the applications in genetic studies. We summarize this work and discuss future plans in Section \ref{Sect.Discuss}. The Appendix contains the proofs of theorems as well as extra numerical  results.

\section{Models of Hypotheses}\label{Sect.Formulation}

In this section we consider two well-connected settings for hypotheses: the Gaussian mean model and generalized linear model. The GMM assumes that the distribution of the vector of $n$ input test statistics $T = (T_1, ..., T_n)$ is joint Gaussian:
\begin{equation}
\label{equ.GMM}
T \sim N(\mu, \Sigma), 
\end{equation}
where the correlation matrix $\Sigma$ is known or can be reliably estimated, the mean vector $\mu$ is the unknown parameter to be tested: 
\begin{equation}
\label{equ.GMM.H0Ha}
H_0: \mu = 0 \quad \text{ versus } \quad H_1: \mu \neq 0.
\end{equation}
This family-wise hypothesis testing problem is also referred as the signal-detection problem \cite{Donoho2004}. The magnitude of a nonzero mean element $\mu_i$ represents an SNR, serving as a measure for signal strength. Certainly, a higher SNR is preferred because it shall lead to higher statistical power. Depending on specific data analysis purpose, the input $p$-values for the gGOF statistics defined in (\ref{equ.gGOFstat}) is either one-sided or two-sided: 
\begin{equation}
\label{equ.GMM.pValue}
\text{ One-sided: } P_i = \bar{\Phi}(T_i); \quad \text{ Two-sided: } P_i = 2\bar{\Phi}(|T_i|),
\end{equation}
where $\bar{\Phi}$ denotes the survival function of $N(0, 1)$. For consistency of the presentation, in this paper the input statistics are always standardized such that $\Sigma$ is a correlation matrix with diagonal of  1's. Such data standardization is consistent with the fact that any point-wise rescaling for $T_i$ won't affect the input $p$-values nor the gGOF statistics. 

The GLM is defined as
\begin{equation}
g(E(Y_k | X_{k\cdot}, Z_{k\cdot})) = X'_{k\cdot}\beta + Z'_{k\cdot} \gamma,
\label{equ.the GLM}
\end{equation}
where for the $k$th subject, $k = 1, ..., N$, $Y_k$ denotes the response value, $X'_{k\cdot} = (X_{k1}, ..., X_{kn})$ denotes the design matrix $X_{N\times n}$'s $k$th row vector of $n$ inquiry covariates, $Z'_{k\cdot} = (Z_{k1}, ..., Z_{km})$ denotes $Z_{N\times m}$'s $k$th row vector of $m$ control covariates.  
For example, in the gene-based SNP-set studies, a gene contains $n$ SNPs (single-nucleotide polymorphisms), $X_{k\cdot}$ is the genotype vector of the $n$ SNPs of the $k$th individual;  $Z_{k\cdot}$ is the vector of $m$ control covariates, such as the intercept and other environmental and genetic variants. The null hypothesis is that conditioning on the control covariates, none of the inquiry covariates are associated with the outcome: 
\begin{equation}
\label{equ.the GLM.H0Ha}
H_0: \beta=0 \text{ vs. } H_1: \beta \neq 0. 
\end{equation}
 The function $g$ is called the link function based on the distribution of $Y_k$ given $X_{k\cdot}$ and $Z_{k\cdot}$. Generally, $Y_k$ follows a distribution in the exponential family with density function 
\[
f(y_k)=\exp\{\frac{y_k\theta_k-b(\theta_k)}{a_k(\phi)}+c(y_k,\phi)\},
\] 
where $\theta$ is the natural parameter, $\phi$ is the dispersion parameter, and the functions $a_k$, $b$ and $c$ are given. 

Through the joint model-fitting, the GLM with correlated covariates is connected to the GMM in (\ref{equ.GMM}). Joint model-fitting means that the estimation of $\beta$ is made simultaneously. Here we consider the classic one-step maximum likelihood estimation (MLE, cf. Theorem 4.19 and Exercise 4.152 of \cite{shaostat}). Let $\mu^{(0)}_k=g^{-1}(Z_{k\cdot}^{\prime} \gamma^{(0)})$, $k=1,...,N$, be an initial estimation, where $\gamma^{(0)}$ is the MLE estimator of $\gamma$ under $H_0$. Denote the estimated diagonal matrix of weights be $W={\rm diag}({\rm Var}^{(0)}(Y_k))_{1 \leq k \leq N}={\rm diag}(a(\phi^{(0)})b^{\prime\prime}(\theta^{(0)}_k))_{1 \leq k \leq N}$, where $\phi^{(0)}$ and $\theta^{(0)}_k$ are also the MLE of $\phi$ and $\theta_k$ under $H_0$, respectively. For example in the logit model, if $Z$ is a column of ones for the intercept, then $\mu^0=\bar{y}$, $\gamma^{(0)}=g(\bar{y})$. Define $\tilde{X}=W^{1/2}X$, $\tilde{Z}=W^{1/2}Z$. $\tilde{H}=\tilde{Z}(\tilde{Z}^\prime \tilde{Z})^{-1}\tilde{Z}^\prime$ is the projection matrix onto the column space of $\tilde{Z}$. Assume $\tilde{X}^\prime(I-\tilde{H})\tilde{X}$ is of full rank, it can be shown (see Theorem \ref{thm.SNR.the GLM}) that the joint model-fitting estimator of $\beta$ is 
\begin{equation}
\label{eq.jointBeta.the GLM}
	\hat{\beta}_J=  (\tilde{X}^\prime(I-\tilde{H})\tilde{X})^{-1}X^\prime(Y-\mu^{(0)}) \overset{D}{\to} N(\beta, (\tilde{X}^\prime(I-\tilde{H})\tilde{X})^{-1}),
\end{equation}
as $N\to \infty$. To obtain the input $p$-values by (\ref{equ.GMM.pValue}) for the gGOF, the input statistics are the standardized $\hat{\beta}_J$: 
\begin{equation}
\label{equ.Tjoint.the GLM}
	T_J  = \tilde{\Lambda}\hat{\beta}_J \overset{D}{\to} N( \tilde{\Lambda}\beta,  \tilde{\Lambda}( \tilde{X}^\prime(I- \tilde{H}) \tilde{X})^{-1} \tilde{\Lambda}),
\end{equation}
where the diagonal matrix $ \tilde{\Lambda} =\text{diag}(\frac{1}{\sqrt{ \tilde{\lambda}_i}})_{1 \leq i \leq n}$, with $ \tilde{\lambda}_i=(( \tilde{X}^\prime(I- \tilde{H}) \tilde{X})^{-1})_{ii}$ being the diagonal elements of $( \tilde{X}^\prime(I- \tilde{H}) \tilde{X})^{-1}$.  

A special case of the GLM is the linear regression model (LM), where $Y_k$ is Gaussian. The model equation becomes 
\begin{equation}
	Y = X\beta + Z \gamma + \epsilon,  
\label{equ.LM}
\end{equation}
where $X_{N\times n}$ and $Z_{N\times n}$ are still the inquiry and controlling design matrices with their $k$th row vectors being  $X_{k\cdot}^\prime$ and $Z_{k\cdot}^\prime$, respectively. The error term  $\epsilon \sim N(0, \sigma^2 I_{N \times N})$, where the variance $\sigma^2$ is assumed known or can be consistently estimated. 
Here the one-step MLE is the same as the least-squares (LS) estimation for the joint model-fitting. In particular, the weights matrix $W={\rm diag}(1/\sigma^2)$; the initial estimation $\mu^{(0)} = \tilde{H}Y$ is the projection of $Y/\sigma$ onto the column space of $Z$, i.e., the LS estimator $\hat{Y}/\sigma$ under $H_0$. Note that the standardized statistics exactly follow the normal distribution: 
\begin{equation}
\label{equ.Tjoint.reg}
	T_J  = \Lambda\hat{\beta}_J/\sigma \sim N(\Lambda\beta/\sigma, \Lambda(X^\prime(I-H)X)^{-1}\Lambda),
\end{equation}
where the diagonal matrix $\Lambda =\text{diag}(\frac{1}{ \sqrt{\lambda_i}})_{1 \leq i \leq n}$, with $\lambda_i=((X^\prime(I-H)X)^{-1})_{ii}$.

From the perspective of $T_J$, the GLM model can be considered as a restricted GMM. In the GMM, $\mu$ and $\Sigma$ are defined independently. In the GLM, even though the ``effects'' $\beta$ are defined separately from data $X$ and $Z$, both the mean vector $\mu_{T_J}$ (which is related to the signals under $H_1 $) and the correlation matrix $\Sigma_{T_J}$ depend on the data correlation structure. Consider the formula in (\ref{equ.Tjoint.the GLM}), $\tilde{X}^\prime(I- \tilde{H}) \tilde{X}$ gives a measure of the covariance among the inquiry covariates conditional on the control covariates.  For example, in the LM when $Z$ contains a single column of 1's, $\tilde{X}^\prime(I- \tilde{H}) \tilde{X} = X'(I-J)X$ (where $J$ is a matrix of $1/n$) is exactly the empirical covariance matrix among the columns of $X$. The connection between the input statistics $T_J$ and the data correlation structure raises two consequences. First, the signal strength depends on data correlation. For example, as to be shown in Section \ref{Sect.ITunderthe GLM}, it is often the case that $\tilde{\lambda}_i > 1$, indicating the signal strength in $T_J$ is actually less than the effect size defined by the nonzero elements of $\beta$. Secondly, a proper linear transformation of the input statistics is related to the data correlation, which we will also discuss further in Section \ref{Sect.transformations}.

\section{The gGOF Family}\label{Sect.gGOF}

In this section we first introduce some important hypothesis-testing procedures as examples of the gGOF. Then we provide analytical $p$-value calculations for the gGOF, based on which a study of rejection boundary for the gGOF procedures is also given. 

\subsection{Examples of the gGOF}

The gGOF family defined in (\ref{equ.gGOFstat}) and (\ref{equ.cross.bound.prob}) is based on a very general functional. It covers a broad range of statistics and testing procedures. For example, the initial Kolmogorov-Smirnov test statistic is a simple case of the gGOF statistic (c.f. \cite{komogorov1933} and \cite{shaostat},  page 447). The KS statistic directly measures the difference between  $x=i/n$ and $y=P_{(i)}$. That is, the corresponding $f$ function is defined by
\[
	 f_{KS^+}(x,y)=x-y.
\]
Jager and Wellner introduced a collection of $\phi$-divergence statistics \citep{jager2007goodness}, which are also based on the supremum of functions with a statistic-defining parameter $s$:
\begin{equation}
\label{equ.phiDiverg}
	\begin{array}{ll}
	f^{\phi}_s(x,y)=\displaystyle\frac{1}{s(1-s)}(1-x^sy^{1-s}-(1-x)^s(1-y)^{1-s})\text{, } s\neq 0,1,\\
	f^{\phi}_1(x,y)=x\log(\displaystyle\frac{x}{y})+(1-x)\log(\frac{1-x}{1-y}) \text{, and } \\
	f^{\phi}_0(x,y)=y\log(\displaystyle\frac{y}{x})+(1-y)\log(\frac{1-y}{1-x}).
	\end{array}
\end{equation}
Since only small $p$-values $P_{(i)} < \frac{i}{n}$ (instead of large $p$-values) indicate signals or effects, it is appropriate to consider the one-sided $\phi$-divergence statistics, which can be simply written as 
\begin{equation}
\label{equ.phiDiverg.oneside}
f_s(x,y) =
	\left\{
		\begin{array}{lr}
			\sqrt{2nf^{\phi}_s(x,y)} & \quad y \leq x ,\\
			-\sqrt{2nf^{\phi}_s(x,y)} & \quad y > x.
		\end{array}
	\right.
\end{equation}	
Thus, the one-sided $\phi$-divergence statistics are a special case of the gGOF and cover the HC statistics $HC^{2004}$ and $HC^{2008}$ \citep{Donoho2004, Donoho2008} for $s=2$ and $-1$, respectively: 
\begin{equation}
\label{eq.HC.statistics}
f_{HC^{2004}} =  \frac{\sqrt{n} (x - y)}{\sqrt{y(1-y)}}; \quad 
f_{HC^{2008}} =  \frac{\sqrt{n} (x - y)}{\sqrt{x(1-x)}}.
\end{equation}
Also, the Berk-Jones (BJ) statistic and the reverse  Berk-Jones statistics \citep{berk1979g} correspond to $s=1$ and $0$, respectively.  

Based on the HC and the BJ, the GHC \citep{barnett2016generalized} and the GBJ \citep{sun2017set} were created to explicitly account for correlations, largely motivated from an interpretation perspective. In particular, the GHC is a version of the HC except the denominator was replaced by the formula of the variance of the numerator, which is related to correlated Bernoulli random variables. Similarly, the GBJ is a version of the BJ except with the probability calculation conditional on the correlated input statistics.  The statistic formulas of the GHC and the GBJ are somewhat complex, but both can be written in the form of the cross-boundary probability of the gGOF in (\ref{equ.cross.bound.prob}).

Furthermore, 
the traditional Bonferroni and the false discovery rate (FDR) procedures also can be considered as special cases in the gGOF family. At a given significance level $\alpha$, the Bonferroni procedure is to reject $H_0$ if any $p$-values are less than $\frac{\alpha}{n}$. Following the formula in (\ref{equ.gGOFstat}), we can consider a gGOF statistic with the $f$ function:
\[
	 f_{B}(x, y)=\frac{\alpha}{n} - y.
\]
Similarly, the FDR procedure is to reject $H_0$ if any $p$-values are less than $\alpha\frac{ i}{n}$, and thus we can consider a gGOF statistic corresponding to 
\[
	 f_{FDR}(x, y)=\alpha x - y.
\]
No truncation is required for both statistics, i.e., $\mathcal{R} = \{1 \leq i \leq n \}$. The Bonferroni procedure and the FDR procedure are equivalent to these two statistics respectively at the decision-making threshold $b=0$.

\subsection{P-value Calculations for the gGOF under Dependence}\label{Sect.Pcalcu}

In this section we propose a strategy to calculate the $p$-values for the gGOF statistics and the adaptive version in (\ref{equ.gGOFstat}) and (\ref{equ.doubleomnibus}). First, Theorem \ref{thm.H0distn.equalCorr} provides an exact calculation under the equal-correlation case. Denote $\phi(z)$ and $\Phi(z)$ the density and the distribution function of $N(0,1)$,  respectively. 

\begin{theorem}
\label{thm.H0distn.equalCorr}
Consider the input statistics $T$ in (\ref{equ.GMM}) with $\mu=0$, $\Sigma_{ij} = \rho$ for all $i\neq j$. 
Assume  $\mathcal{R} = \{i:  k_0 \leq i \leq k_1 \}$. Let $U_{(1)} \leq ... \leq U_{(n)}$ be the order statistics from $n$ {\it iid} Uniform$(0,1)$ random variables. For any   gGOF statistic in (\ref{equ.gGOFstat}): 
\begin{equation}
\label{equ.H0distn.equalCorr}
\mathbb{P}(S_{n, f, \mathcal{R}} < b)=\int_{-\infty}^\infty \mathbb{P}(U_{(i)}>c_{i}(z),i=k_0,...,k_1) \phi(z) dz, 
\end{equation}
where for the one-sided input $p$-values in (\ref{equ.GMM.pValue}),
\[
c_{i}(z)=1-\Phi\left(\frac{\Phi^{-1}(1-u_i)-\sqrt{\rho}z}{\sqrt{1-\rho}}\right),
\]
with $u_i$ given in (\ref{equ.cross.bound.prob}), and for the two-sided input $p$-values,
\[
c_{i}(z)=1- \Phi\left(\frac{\Phi^{-1}(1-u_i/2)-\sqrt{\rho}z}{\sqrt{1-\rho}}\right) + \Phi\left(\frac{-\Phi^{-1}(1-u_i/2)-\sqrt{\rho}z}{\sqrt{1-\rho}}\right).
\]
\end{theorem}

Efficient calculation methods for the exact or the approximate value of $\mathbb{P}(U_{(i)}> a_i, i=k_0,...,k_1)$ have been given in \cite{Zhang2016distributions}. The integration over $z$ can be easily calculated by numerical methods. 

In real data analysis, the structure of the correlation matrix $\Sigma$ could be complex. To address this situation, we provide two approaches to approximate the distributions of the gGOF statistics based on the exact calculation developed above. 

The first approach is called the weighted average method (WAM). It is specially designed for correlation matrices that are roughly Toeplitz. That is, the off-diagonal lines have equal elements, i.e., $\Sigma(l, k)=\rho_j, j = {|l-k|}$. This assumption is appropriate for data with proportionally decaying correlations. For example, in time series data or genetic data, correlations often decay over time or locations. Let $G_{\Sigma}(b) \equiv \mathbb{P}(S_{n, f, \mathcal{R}} < b | \Sigma)$ be the distribution function of $S_{n, f, \mathcal{R}}$, where $\Sigma$ is the correlation matrix of the input statistics $T$. Denote $\Sigma_j$ the equal-correlation matrix with correlation $\rho_j$. The WAM approximation is 
\begin{equation*}
G_{\Sigma}(b) = \sum_{j=1}^{\alpha n}\omega_j G_{\Sigma_j}(b).
\end{equation*}
Theoretical \citep{cai2010optimal} as well as our empirical studies show that the near off-diagonal components are more important for characterizing $\Sigma$. Thus, we propose a band-width parameter $\alpha \in (0, 1)$ to truncate the far-end off-diagonals. The weights are based on the relative size of the off-diagonal lines: $\omega_j = \frac{n-j}{((1+\alpha)n-1)(1-\alpha)n/2}$, $j=1, ... ,\alpha n$. Empirical results show that $\alpha=0.5$ is a robust choice in most cases. When $\Sigma$ is not exactly Toeplitz, we can define $\rho_j$ be the average correlation on the $j$th off-diagonal of $\Sigma$.

When the correlation matrix $\Sigma$ is more complicated than Toeplitz, we propose a second approach based on the LOESS (LOcal regrESSion).  That is, we obtain the $G_{\Sigma}(b)$ curve by a local-smoothing curve fitting of $G_{\Sigma_j}(b)$ around the point $b$. $\Sigma_j$ still represents the equal-correlation matrix with constant correlation $\rho_j$. However, instead of focusing on the near off-diagonals in WAM, here $\rho_j$ is chosen from all elements in $\Sigma$.  The idea is that large correlation elements in $\Sigma$, even if not necessarily close to the diagonal, could likely have non-negligible influence. The algorithm steps are given below:
\begin{enumerate}
\item[(1)] Set $m$ equal-distance neighbor points in the interval around $b$, i.e., $b_i \in [b-\epsilon, b+\epsilon]$, $i=1,...,m$. 
\item[(2)] For each $b_i$, randomly choose $N$ off-diagonal elements $\rho_j$ of $\Sigma$ to calculate $y_{ij}=G_{\Sigma_j}(b_i)$, $j=1,...,N$.  
\item[(3)] Use $y_{ij}$ as the input data for a local polynomial regression with tri-cube weights to predict the curve $G_{\Sigma}(b)$ \citep{cleveland1992local}. 
\end{enumerate}
Note that the random choice in step (2) allows a higher chance of picking correlations from near-diagonal elements of $\Sigma$, and thus essentially gives a heavier weight for them. For implementation, we found $m=10$, $\epsilon=1$, $N=n$, and the curve function of quadratic polynomials often provide a good rule of thumb. Comparing these two approaches, WAM is often more accurate when $\Sigma$ is close to Toeplitz pattern, while LOESS is often more robust when correlation is more complex.  

Regarding the $p$-value calculation of the digGOF, as described in (\ref{equ.cross.bound.prob.omni}) the calculation is still about the cross-boundary probability. Thus, it follows the same methods given above. The key is to obtain the boundaries $u^\star_i$. For efficient computation, we consider double-adaptation over a discrete sequence of functions and truncations $\{(f_1, \mathcal{R}_1), (f_2, \mathcal{R}_2), ... \}$. Here any monotone functions $f_j$ would work. For example, one choice could be the $\phi$-divergence family in (\ref{equ.phiDiverg.oneside}), where all $f_s$ functions $s \in [-1, 2]$ ensure the asymptotic optimality for detecting weak-and-rare signals \citep{jager2007goodness}. At the same time, we consider truncation domains $\mathcal{R}_j = \{i: k_{0j} \leq i \leq k_{1j}\}$ based on the index $i$ only. Note that for a given data, the adaptation of $i$ to $\mathcal{R}_j$ for all $k_0 \leq k_1 \in \{1, ..., n\}$ is equivalent to the adaptation of $P_{(i)}$ to $\{\alpha_{0j} \leq P_{(i)} \leq \alpha_{0j} \}$ for all $\alpha_{0j} \leq \alpha_{1j} \in [0, 1]$. Moreover, computation can be significantly simplified. Specifically, denote the digGOF statistic in (\ref{equ.doubleomnibus}) as $S_o =  \inf_j G_j(S_j)$, its survival function under $H_0$ is 
\begin{align*}
\mathbb{P}(S_o>s_o) =\mathbb{P}(S_j(P_1,...,P_n)<G_j^{-1}(s_o), \text{ for all } j) = \mathbb{P}(P_{(1)}>u^\star_1(s_o),...,P_{(n)}>u^\star_n(s_o)),
\end{align*}
where $u^\star_i(s_o)=\sup_j u_{ji}(s_o)$. Here, for each given $j$, we only need to calculate $u_{ji}$ within $k_{0j} \leq i \leq k_{1j}$ because 
\begin{equation}
u_{ji}(s_o) = 
\begin{cases}
f^{-1}_j(\frac{i}{n}, G_j^{-1}(s_o)) & \text{ if } k_{0j} \leq i \leq k_{1j}, \\
0 & \text{ otherwise. }
\end{cases}
\end{equation}
When the input statistics are {\it iid}, or they are under the equal-correlation, the calculation can be exact based on \cite{Zhang2016distributions} or Theorem \ref{thm.H0distn.equalCorr} above. Under complex correlations, approximation algorithms by WAM or LOESS will be applied. 

\subsection{ Rejection Boundary Analysis} 

For any   gGOF statistic, let $b_\alpha$ be the threshold in (\ref{equ.cross.bound.prob}) that controls the type I error rate at $\alpha$. For this statistic to accept $H_0$, the acceptance region is 
\begin{equation}
\mathcal{A}_{S} = \{P_{(i)}: P_{(i)}>u_i(b_\alpha) \text{, for all } i \text{ and } P_{(i)} \in \mathcal{R} \}.
\label{equ. A}
\end{equation}
The rejection region is $\mathcal{A}^c_{S}$ because $H_0$ is rejected whenever $P_{(i)} \leq u_i(b_\alpha)$ at any $i$.  Thus, for any   gGOF statistic at given $n$, $\Sigma$ and $\alpha$, we call the series $\{u_1(b_\alpha), ..., u_n(b_\alpha) \}$ the rejection boundary (RB) of this statistic.  A uniformly higher RB curve indicates a uniformly higher statistical power over all possible signal strength. However, when RBs cross, the relative power advantages depend on signal patterns. 

\begin{figure}[!t] 
	\begin{center}
		\subfloat{\includegraphics[width=2.5in,height=2.5in]{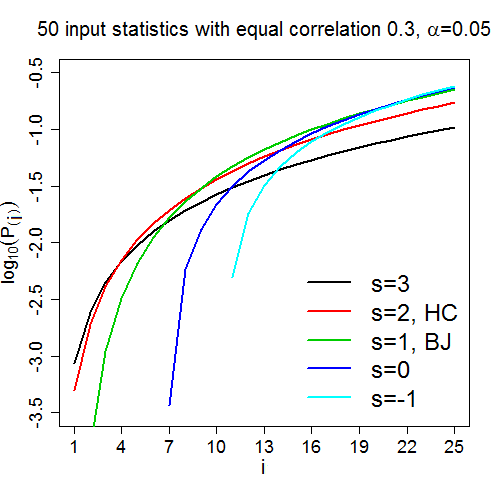}}
		\subfloat{\includegraphics[width=2.5in,height=2.5in]{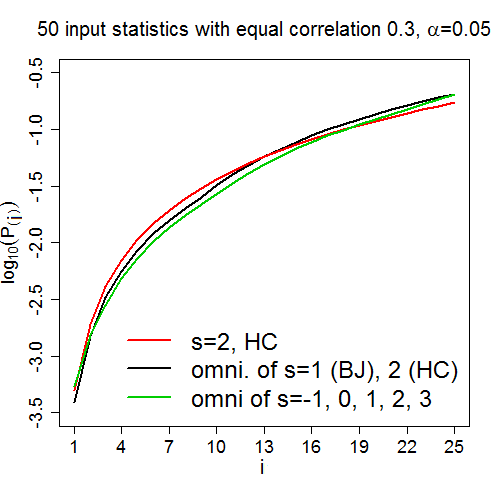}}
	\end{center}
	\caption{The log rejection boundary of the $\phi$-divergence statistics and their omnibus versions. The null hypothesis is rejected as long as one value of $\log_{10}(P_{(i)})$ is below the boundary.  
	Left panel: RBs for $\phi$-divergence statistics with $s=3$ (black), $2$ (i.e., $HC^{2004}$, red), $1$ (i.e., the BJ, green), $0$ (i.e., reverse BJ, blue) and $-1$ (i.e., $HC^{2008}$ or reverse HC, cyan). Right panel: RBs for $HC^{2004}$ (red), omnibus over $HC^{2004}$ and the BJ (black), and omnibus over $\phi$-divergence statistics with $s \in \{-1,0,1,2,3\}$ (green).}
\label{fig:omnibound}
\end{figure}

Figure~\ref{fig:omnibound} illustrates the RBs at the logarithm scale for various gGOF statistics. The left panel evaluates various $\phi$-divergence statistics with $s \in \{ -2, -1, 0, 1, 2, 3\}$. The RB curves show an interesting pattern: statistics with larger $s$ (e.g., $HC^{2004}$ with $s=2$) have higher RBs at the top ranked $p$-values. If signals are strong and sparse so that they correspond to the smallest input $p$-values, then these statistics are more sensitive to these signals and thus likely give more statistical powerful. 
On the other hand, statistics with small $s$ are more sensitive to the lower ranked $p$-values, indicating their advantages to detect signals that are weaker and/or denser. Such an observation also indicates that a proper $p$-value truncation domain $\mathcal{R}$ could benefits the statistic power. 
As shown in the right panel, the omnibus methods over various $f_s$ functions provides a more balanced and thus more robust solution. The RB curves of the omnibus tests tend to be closer to the best statistic at all positions.

The study of RB is a good way to understand the performance of the gGOF statistics. However, it cannot replace the study of statistical power. In particular, the distances between two RB curves at different locations do not proportionally reflect their relative advantages in terms of statistical power. That is, depending on signal and correlation patterns, a small RB curve difference at some locations could be more important than a larger RB curve difference at other locations. Thus, if two RBs cross, it is hard to say which method is more powerful, especially when the signal pattern is to be determined in a real data analysis. 

\section{Data Transformations}\label{Sect.transformations}

In this section we consider linear transformations of the input statistics before getting the input $p$-values. Proper transformation could utilize correlations to increase signal strength and thus improve the group-testing procedures.  The first part of this section addresses sparse signals; the second part studies arbitrary dense signals and correlations under a bivariate regression model. 

\subsection{Sparse Signals and Transformations}\label{Sect.IT.sparse}

For sparse signals, we first illustrate the problem under the GMM, and then extend the results into the GLM setting. 

\subsubsection{Transformations Under the GMM}

For a vector of input test statistics under the GMM in (\ref{equ.GMM}), \cite{Hall2010} proposed the de-correlation transformation (DT) and the innovated transformation (IT). Define the Cholesky factorization of $\Sigma$ by a lower triangular matrix $Q$, i.e. $\Sigma = QQ'$ or $Q'\Sigma^{-1}Q = I$. Define $U = Q^{-1}$, which is also a lower triangular matrix. We have $\Sigma^{-1} = U'U$ or $U\Sigma U' = I$. 

The DT is simply to apply $U$ transformation such that the input statistics become independent:
\begin{equation}
\label{equ.DT.stat}
T^{DT} = U T \sim N(U\mu, I). 
\end{equation}
For the IT, we define it in a more specific way as below: 
\begin{definition}
\label{def.IT}
For a Gaussian random vector $T$ in (\ref{equ.GMM}), if the locations of the nonzero elements of $\mu$ do not depend on $\Sigma$, then the IT of $T$ is 
\begin{equation}
\label{equ.IT.stat}
T^{IT} = D\Sigma^{-1}T \sim N(D\Sigma^{-1}\mu, D\Sigma^{-1}D),
\end{equation}
where $D = {\rm diag}(1/\sqrt{(\Sigma^{-1})_{ii}}, i = 1, ..., n)$. 
\end{definition}
The rescaling matrix $D$ is for making sure the variances of $T^{IT}$ are 1, to be consistent with the GMM definition in (\ref{equ.GMM}). Thus the nonzero means in $D\Sigma^{-1}\mu$ represent the SNRs. A subtle but more important requirement is that the locations of the nonzero means are independent of the correlation. This requirement rules out the inverse transformation procedure because the locations of the nonzero elements of $\Sigma^{-1}\mu$ are determined not only by $\mu$ but also by $\Sigma^{-1}$. In other words, this requirement anchors a baseline for the IT procedure and make it a directional transformation. Therefore, the definition of the IT is in consistence with the direction of SNR change.  

\begin{figure}[!t] 
	\begin{center}
		\subfloat{\includegraphics[width=2.5in,height=2.5in]{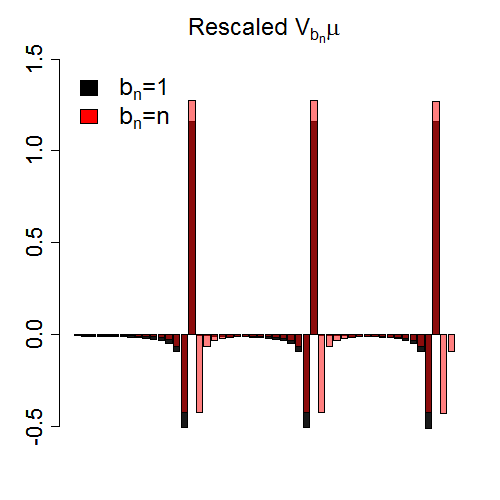}}
		\subfloat{\includegraphics[width=2.5in,height=2.5in]{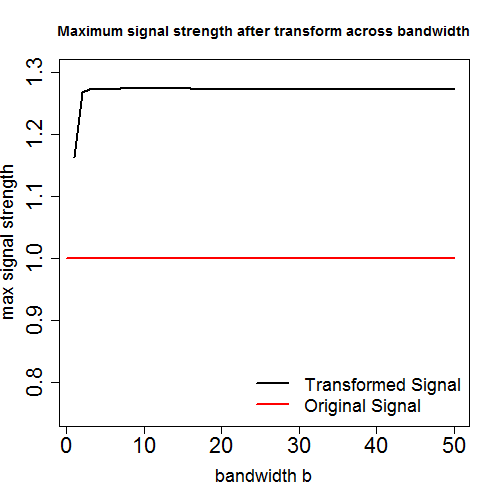}}\\
		\subfloat{\includegraphics[width=2.5in,height=2.5in]{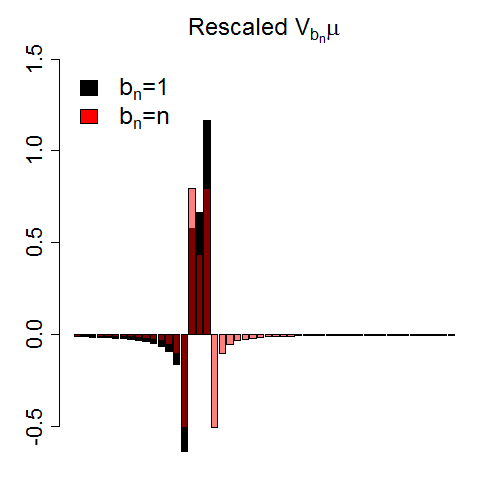}}
		\subfloat{\includegraphics[width=2.5in,height=2.5in]{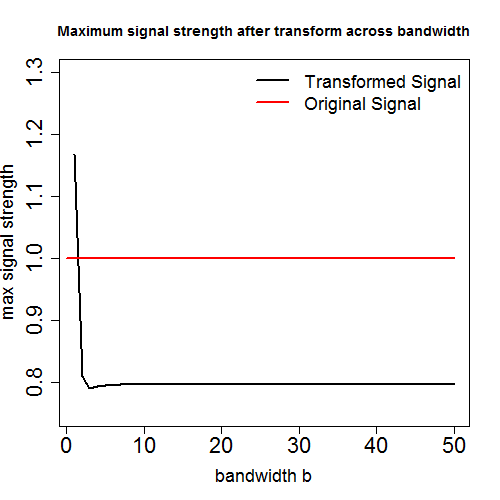}}
	\end{center}
	\caption{SNR before and after transformations. First row: $\mu_{16}=\mu_{32}=\mu_{48}=1$ and other $\mu_i=0$. Second row: $\mu_{16}=\mu_{17}=\mu_{18}=1$ and other $\mu_i=0$. Bars on left: all SNRs after the DT (black) and the IT (red) over positions $i = 1, ..., 50$. Curves on right: max SNRs over $b_n = 1, ..., n$. $\Sigma$ is polynomial-decay with $\Sigma_{ij}=|i-j|^{-1}$, $i\neq j$ and $\Sigma_{ii}=1$, $i=1,...,50$.  
	}
\label{fig:trans_sig}
\end{figure}

It should be noted that the transformations proposed by \cite{Hall2010} are broader:
\begin{equation}
V_{b_n} T \sim N(V_{b_n}\mu, V_{b_n} \Sigma V^\prime_{b_n}),
\label{equ.innovTrans}
\end{equation} 
where $V_{b_n}$ is a transforming matrix with some off-diagonals being truncated and all columns being normalized (which is also for rescaling the statistics). The band width of the non-truncated diagonals is controlled by $b_n$: when $b_n=1$, $V_1 = U$, the transformation is the DT; when  $b_n=n$, $V_1 = D\Sigma^{-1}$, the transformation is the IT. Besides providing a spectrum of transformations, introducing $b_n$ also provides technique convenience in theoretical proof of asymptotic optimality as $n\to\infty$ (also see Theorem \ref{thm.iHC.optimality.GLM} below). Meanwhile, in practical data analyses with finite $n$, the DT and the IT are often enough to represent the two extremes of the signal strength after transformation. Figure \ref{fig:trans_sig} demonstrates a study on the influence of $b_n$. As demonstrated in the first row, when the signals are separated far from each other, the IT is often better than the DT in terms of a larger maximum SNR. When signals are clustered (second row) the transformed SNRs depend more on the off-diagonal correlations. Here the inverse correlation matrix has negative entries, which leads to signal cancellations in IT, and the DT is a better choice. The performance of other transformations at different $b_n$ values are between the IT and the DT. For simplicity, in practical data analysis we could focus on choosing from these two transformations. 

In the following, we study the conditions for the DT and the IT to increase SNRs. One scenario is illustrated in the first row of Figure \ref{fig:trans_sig}. Roughly speaking, the signals are sparse relatively to the correlations.  It is also referred as the sparse-signal-sparse-correlation condition in literature \citep{fan2013optimal}. However, we emphasize that the relativeness of the sparsity. That is, as long as signals are sparse enough, the correlation could be dense. Also note that the condition is irrelevant to positive or negative correlations. 

Specifically, the reasoning is based on a few important linear algebra properties about $Q$, $U$ and $\Sigma^{-1}$:
\begin{enumerate}
\item Since $\Sigma$ is a symmetric positive definite matrix with diagonal of 1's, the diagonal of $Q$ can be calculated as $Q_{j,j}=\sqrt{\Sigma_{j,j}-\sum_{k}Q^2_{j,k}}=\sqrt{1-\sum_{k}Q^2_{j,k}} \leq1$. The diagonal elements of $U=Q^{-1}$ satisfy $U_{j,j}=1/Q_{j,j} \geq 1$.

\item $\Sigma^{-1}$ is symmetric with diagonal elements 
\[
(\Sigma^{-1})_{j, j} = \sum_{k=0}^{n-j} u^2_{j+k, j}, 
\]
with special cases:  $(\Sigma^{-1})_{1, 1} = \sum_{k=0}^{n-1} u^2_{1+k, 1}$ and $(\Sigma^{-1})_{n, n} = u^2_{n, n}$. Thus the diagonal elements $U_{j,j}$ are increasing in $i$, with 
\begin{equation}
\label{equ.U.SigmaInverse.diagonal}
1 = U_{1,1} < U_{2,2} < ... <U_{n,n} = \sqrt{(\Sigma^{-1})_{n,n}}, 
\end{equation}
and $(\Sigma^{-1})_{1,1} = ... =  (\Sigma^{-1})_{n,n} > 1$. 

\item If $\Sigma$ is polynomially decaying, then $Q$, $U$ and $\Sigma^{-1}$ are all polynomially decaying \citep{sun2005wiener, Hall2010}. 

\end{enumerate}

To understand the idea easily, we can first start with the simplest case where there is only one non-zero element of $\mu$ at position $j$, i.e., $\mu_{j}=A$ for a constant $A$ and $\mu_{i}=0$ for all $i \neq j$. For the DT, by (\ref{equ.U.SigmaInverse.diagonal}), the post-transformed mean vector $U\mu$ has an increase SNR at the $j$th element $(U\mu)_j = AU_{j,j} \geq A$. Similarly,  the SNR after the IT is even further increased over the DT because $(D\Sigma^{-1}\mu)_j =   \frac{1}{\sqrt{(\Sigma^{-1})_{j,j}}} (\Sigma^{-1})_{j,j} A = A\sqrt{(\Sigma^{-1})_{j,j}}  \geq AU_{j,j}$. 

\begin{figure}[!t] 
	\begin{center}
		\subfloat{\includegraphics[width=3in,height=3in]{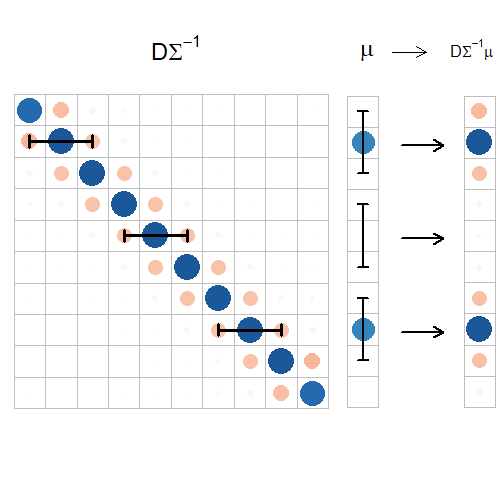}}
	\end{center}
	\caption{Demonstration of the IT when signals (represented by dots in $\mu$) are sparse and widely spread relatively to the non-negligible correlations (represented by dots in $D\Sigma^{-1}$). Empty cells represent zero or close-to-zero elements. 
	}
\label{fig:trans_diagram}
\end{figure}

The deduction can be extended to a more general case of multiple true signals, which is demonstrated by the diagram in Figure \ref{fig:trans_diagram}. Dots in $\mu$ represent nonzero elements; segments in $D\Sigma^{-1}$ represent the band width of non-negligible correlations. Because diagonal elements of $D\Sigma^{-1}$ are larger than 1, if the distances between signals are wider than the correlation band, i.e., the signals are sparser than the correlations, after transformation the SNR in $D\Sigma^{-1}\mu$ are increased.  The off-diagonal correlations could create more nonzero elements in $D\Sigma^{-1}\mu$, but they are likely smaller, and thus are less influential. On the other hand, the inverse transformation from $D\Sigma^{-1}\mu$ back to $\mu$ will reduce SNR. But because nonzero locations in $D\Sigma^{-1}\mu$ depend on the correlation $\Sigma$, by Definition \ref{def.IT} we do not consider this inverse procedure as the IT. 

The idea of Figure \ref{fig:trans_diagram} has been mathematically formulated under the GMM \citep{Arias2010, Hall2010}. Specifically, let $\mu_{j}=A$ for $j\in M^{\ast }=\left\{j_{1},...,j_{K}\right\}$, where $M^{\ast }$ represents the locations of true signals. A sufficient condition for the DT and the IT to increase SNR is: 1) the signals are sparse, e.g., $K = o(n)$; 2) the signal locations $M^{\ast }$ is uniformly distributed over $\{1, ..., n\}$; and 3) correlation matrix $\Sigma$ is polynomially decaying, such that $U$, $Q$, and $\Sigma^{-1}$ are also polynomially decaying. These requirements guarantee that the signals are widely spread with high chance, are sparse relatively to correlations, and are independent of the correlation structure. Under this condition, after transformation, the SNRs are still roughly $AU_{j,j}$ for the DT and $A\sqrt{(\Sigma^{-1})_{j,j}}$ for the IT (cf. Lemma 6 in \cite{Arias2010} and Lemma 11.2 in \cite{Hall2010}). Thus  the DT and the IT still strengthen the signals. Below we extend the study to the GLM.

\subsubsection{Transformations Under the GLM}\label{Sect.ITunderthe GLM}

In this section we study the transformations under the GLM in (\ref{equ.the GLM}). First we consider the special case of linear regression model in (\ref{equ.LM}). Theorem \ref{thm.SNR.LM} relates the transformation types to the model-fitting types. That is, we can consider the marginal model-fitting as the IT of the joint model-fitting. When signals are sparse, no matter what the correlation structure of the data is, the marginal model-fitting is better than the de-correlation strategy, the latter is better than the joint model-fitting. 

\begin{theorem}
\label{thm.SNR.LM}
Consider linear regression model in (\ref{equ.LM}) with the error term $\epsilon \sim N(0, \sigma^2 I)$, where $\sigma$ is assumed known. $H = Z(Z'Z)^{-1}Z'$ is the projection matrix onto the columns of $Z$. Denote $X_j$ the $j$th column of $X$, $j=1, ..., n$. 
\begin{enumerate}
\item Under the least squares estimation (assuming it exists), the test statistics by the joint model-fitting are $T_J$ in (\ref{equ.Tjoint.reg}), the test statistics by the marginal model-fitting are 
\begin{equation}
\label{equ.Tmarg}
T_M = CX'(I-H)Y/\sigma \sim N(\Sigma_{T_M} C^{-1}\beta/\sigma, \Sigma_{T_M}),
\end{equation}
where $C = {\rm diag}( \frac{1}{\sqrt{X'_j(I-H)X_j}})_{1 \leq j \leq n}$, $\Sigma_{T_M} = CX'(I-H)XC$ is the correlation matrix of $T_M$. 

\item The IT of $T_J$ is $T_M$, i.e., $T_J^{IT} = T_M$. The DTs of $T_J$ and $T_M$ are the same:
\[
T_J^{DT} = T_M^{DT} = U_M C X'(I-H)Y / \sigma \sim N( (C U'_M)^{-1}\beta / \sigma, I),
\]
where $U_M$ is the inverse of the Cholesky factorization of $\Sigma_{T_M}$, i.e., $U_M \Sigma_{T_M} U'_M = I$. 

\item If under $H_1$ $\beta$ has one nonzero element $A > 0$, then $E(T_J)  \leq E(T_J^{DT})  \leq E(T_M)$. 
\end{enumerate}
\end{theorem}

A few remarks can be made. First, $T_J$ and $T_M$ can be mutually transformed by multiplying the inverse correlation matrix. However, by definition we do not say $T_J$ is the IT of $T_M$ because the nonzero locations in $E(T_M)$ depend on the correlations.  Treating the joint model-fitting as the starting point is reasonable also because the estimate $\hat{\beta}_J$ is unbiased. The marginal estimator $\hat{\beta}_M$ is biased due to correlations, but it could provide more statistical power for the signal detection problem. Secondly, the correlations among input statistics in $T_M$ is consistent with the correlations among covariates, while the correlations among input statistics in $T_J$ are the inverse. For example, if the covariates are positively correlated, then the starting point $T_J$ often has negative correlations. Therefore, $T_M$ is the IT of $T_J$ transformed by a matrix of positive correlations. Thirdly, the Theorem \ref{thm.SNR.LM} considers one signal for simplicity, but the similar result can be shown as long as the signals are sparser than the correlation. This situation is illustrated in Figure \ref{fig:trans_diagram} and the formal result is given in Theorem \ref{thm.iHC.optimality.GLM}.  
Fourthly, Theorem \ref{thm.SNR.LM} assumes the LSE of the joint model-fitting exists, which is satisfied if $X'(I-H)X$ is full rank. Even if it does not exist (e.g., high dimensional data analysis problem with $N < n+m$), the marginal fitting is still a good choice due to it's simple computation and it's connection with the IT. 

The result of Theorem \ref{thm.SNR.LM} can be extended to the GLM, a more general tool for data analysis. The main difference is that under the GLM, the distributions of $T_J$ and $T_M$ can only be guaranteed asymptotically normal. 

\begin{theorem}
\label{thm.SNR.the GLM}
Consider the GLM in (\ref{equ.the GLM}). 
\begin{enumerate}
\item Under the one-step maximum likelihood estimation (assuming it exists), the test statistics by the joint model-fitting are $T_J$ in (\ref{equ.Tjoint.the GLM}). The test statistics by the marginal model-fitting are 
\begin{equation}
\label{equ.Tmarg_the GLM}
T_M = CX^\prime (Y-\mu^{(0)}) \overset{D}{\to} N(\Sigma_{T_M} C^{-1}\beta, \Sigma_{T_M}),
\end{equation}
where $\mu^{(0)}$ is the MLE estimator of the mean under $H_0$, $C = {\rm diag}( \frac{1}{\sqrt{\tilde{X}^\prime_j(I-\tilde{H})\tilde{X}_j}})_{1 \leq j \leq n}$, $\tilde{X}_j$ is the $j$th column of matrix $\tilde{X}=W^{1/2}X$, and $W={\rm diag}({\rm Var}^{(0)}(Y_k))_{1 \leq k \leq N}$ is the estimated weights matrix under $H_0$. $\Sigma_{T_M} = C\tilde{X}^\prime(I-\tilde{H})\tilde{X}C$ is the correlation matrix of $T_M$,  where $\tilde{H}=\tilde{Z}(\tilde{Z}^\prime \tilde{Z})^{-1}\tilde{Z}^\prime$ and $\tilde{Z}=W^{1/2}Z$. 

\item The IT of $T_J$ is $T_M$, i.e., $T_J^{IT} = T_M$. The DTs of $T_J$ and $T_M$ are the same:
\[
T_J^{DT} = T_M^{DT} = U_M C X^\prime (Y-\mu^{(0)}) \overset{D}{\to} N( (C U'_M)^{-1}\beta , I),
\]
where $U_M$ is the inverse of the Cholesky factorization of $\Sigma_{T_M}$, i.e., $U_M \Sigma_{T_M} U'_M = I$. 

\item If under $H_1 $ $\beta$ has one nonzero element $A > 0$, then $E(T_J)  \leq E(T_J^{DT})  \leq E(T_M)$. 
\end{enumerate}
\end{theorem}

Under the GLM with sparse coefficients $\beta$ and Toeplitz correlation of covariates, we can give the detection boundary of all statistics and show the optimality of iHC for detecting asymptotically weak-and-rare signals. This result readily follows Theorem \ref{thm.SNR.the GLM} and the work of \cite{Hall2010}. Specifically, the technique assumptions are 
\begin{enumerate}
\item[(A)] For $j_k \in M^* = \{j_1, ..., j_K\}$, $\beta_{j} = A_{nj} = \sqrt{2r_j \log n}$ for $j \in M^*$, and $\beta_j= 0$ otherwise. The $\beta$ domain $M^*$ has a size $K=n^{1-\alpha}$, $\alpha\in(1/2, 1)$, indicating sparse signals. Moreover, $M^*$ uniformly distributes on $\{1, ..., n\}$ with equal probability $\binom{n}{K}^{-1}$. 

\item[(B)] Assume that $\Sigma = \tilde{\Lambda}( \tilde{X}^\prime(I- \tilde{H}) \tilde{X})^{-1} \tilde{\Lambda}$ in (\ref{equ.Tjoint.the GLM}) can be written as a Toeplitz matrix that is generated by a spectral density $f$ on $(-\pi, \pi)$, i.e., 
\[
\Sigma_{jk} = \frac{1}{2\pi} \int_{-\pi}^\pi f(t) e^{-i |j-k|t} dt 
\]
is the $|j-k|$th Fourier coefficient of $f$. 

\item[(C)] Assume there exist constants $\gamma > 1$ and $M_0 = M_0(f) > 0$ such that 
\[
|\Sigma_{jk} | \leq \frac{M_0 (f)}{(1 + |j - k|)^{\gamma}},
\]
which indicates that $\Sigma$'s off-diagonals decay at a polynomial rate or faster. 
\end{enumerate}

\begin{theorem}
\label{thm.iHC.optimality.GLM}
Consider the GLM in (\ref{equ.the GLM}). Under the above described assumptions (A)-(C), the detection boundary for $\beta$ is 
\[
\rho_j^*(\alpha) = \frac{ \rho(\alpha)}{\tilde{X}'_j(I-\tilde{H})\tilde{X}_j} \text{, where } 
\rho(\alpha) = 
\begin{cases}
\alpha - 1/2 & 1/2 < \alpha \leq 3/4, \\
(1 - \sqrt{1- \alpha})^2 & 3/4 < \alpha < 1. 
\end{cases}
\]
When $r_j < \rho_j^*(\alpha)$, all hypothesis tests are asymptotically 0 power in the sense that the sum of the type I and the type II error rates converges to 1 as $N>n \to \infty$. When $r_j >\rho_j^*(\alpha)$, if we transform $T_J$ by (\ref{equ.innovTrans}) with the bandwidth $b_n = \log n$, apply the input $p$-values in (\ref{equ.GMM.pValue}) to $HC^{2004}$ statistic in (\ref{eq.HC.statistics}) with $\mathcal{R} = \{1/n \leq P_{(i)} \leq 1/2\}$, and reject $H_0$ whenever $HC^{2004} \geq (\log n)^{5/2}$, then such iHC procedure is asymptotically powerful in the sense that at any fixed type I error rate its power converges to 1 as $N > n\to \infty$.
\end{theorem}

\subsection{Dense Signals and Transformations}\label{Sect.DenseSignals}

Above results show that the marginal model-fitting, which corresponds to the IT, is always a good choice under sparse signals. For more general signal- and correlation-patterns, in this section we compare different testing strategies by the bivariate regression model of two correlated covariates: $X = (X_1, X_2)$. We compare both the individual as well as the summational statistics under both the joint and the marginal model-fitting methods. The summational statistic is a classic strategy for combining test statistics or $p$-values after the Z-transformation \citep{Stouffer1949, littell1971asymptotic}. Under regression model, the distributions of statistics are exactly Gaussian following Theorem \ref{thm.SNR.LM}. The results can also be extended to the GLM following Theorem \ref{thm.SNR.the GLM}. 

Under the bivariate regression model, by (\ref{equ.Tmarg}), we assume the scaled coefficients (representing the sizes of effects) and the correlation of the data are, respectively,
\begin{align*}
C^{-1}\beta/\sigma = \binom{a}{b} \text{ and }
\Sigma_{T_M} = CX'(I-H)XC =
\begin{pmatrix}
1 & \rho \\
\rho & 1
\end{pmatrix}. 
\end{align*}
We have 
\begin{align*}
\Sigma_{T_M}^{-1} = \frac{1}{1-\rho^2}
\begin{pmatrix}
1 & -\rho \\
-\rho & 1
\end{pmatrix}; \quad
U_M = \frac{1}{\sqrt{1-\rho^2}}
\begin{pmatrix}
\sqrt{1-\rho^2} & 0 \\
-\rho & 1
\end{pmatrix}.  
\end{align*}
The two statistics in the marginal-fitting are, 
\[
T_M =  \binom{T_{M1}}{T_{M2}} \sim N(\binom{a+\rho b}{\rho a + b}, \begin{pmatrix}
1 & \rho \\
\rho & 1
\end{pmatrix}).
\]
The rescaled summational statistic under the marginal-fitting is 
\[
T_{MS} = \frac{T_{M1} + T_{M2}}{\sqrt{2(1+\rho)}} \sim N((a+b)\sqrt{\frac{1+\rho}{2}}, \quad 1). 
\]
 For the joint model-fitting, by (\ref{equ.Tjoint.reg}) the statistics are
\[
T_J =\binom{T_{J1}}{T_{J2}} \sim N(\sqrt{1-\rho^2}\binom{a}{b}, \quad 
\begin{pmatrix}
1 & -\rho \\
-\rho & 1
\end{pmatrix}
).
\] 
As shown in Theorem \ref{thm.SNR.LM}, $T_M$ is the IT of $T_J$.  
The rescaled summational statistic under the joint fitting is
\[
T_{JS} = \frac{T_{J1} + T_{J2}}{\sqrt{2(1-\rho)}} \sim N((a+b)\sqrt{\frac{1+\rho}{2}}, \quad 1). 
\]
Interestingly, $T_{JS}$ and $T_{MS}$ follow the same distribution. Therefore, the two model-fitting strategies do not influence the statistical power of such summational statistic. 

\begin{figure}[!t] 
	\begin{center}
		\subfloat{\includegraphics[width=2.1in,height=2.1in]{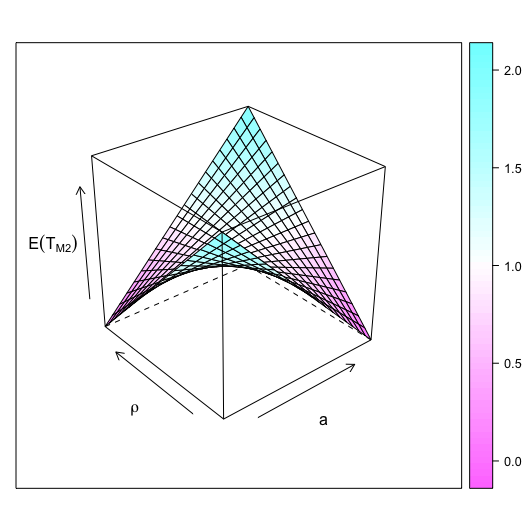}}
		\subfloat{\includegraphics[width=2.1in,height=2.1in]{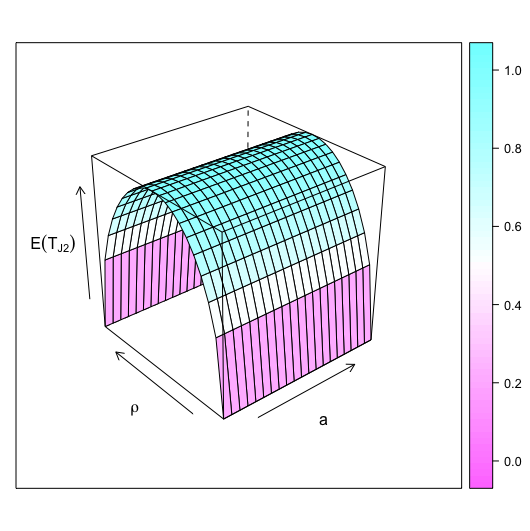}}
		\subfloat{\includegraphics[width=2.1in,height=2.1in]{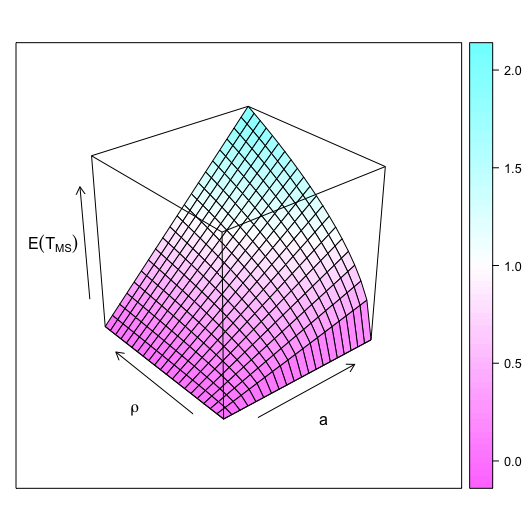}}
	\end{center}
	\caption{SNRs for marginal- and joint-fitting statistics and the summation statistic. Left: $E(T_{M2})$; Middle: $E(T_{J2})$; Right: $E(T_{MS})$. X-axis: $a \in [-1, 1]$; y-axis: $\rho \in [-1, 1]$. Fixing $b=1$.}
\label{fig:3Dsurface_ETs}
\end{figure}

After the DT we have
\[
T_M^{DT} = U_M T_M \sim N( \binom{a+\rho b}{b\sqrt{1-\rho^2}}, I). 
\]
Since $T_{M1}^{DT} = T_{M1}$ and $T_{M2}^{DT} = T_{J2}$, we don't have to consider $T_M^{DT}$ separately. For the other three statistics, due to the symmetry of the expressions we can assume $0 \leq |a| \leq b$ without loss of generality. Under this assumption, $E(T_{M1}) \leq E(T_{M2})$ and $E(T_{J1}) \leq E(T_{J2})$. Since a bigger SNR is more relevant to statistical power of the gGOF, we can simplify the comparisons among three SNRs:
\begin{subequations}
 \label{equ.SNR.2D}
 \begin{align}
E(T_{M2}) &= \rho a + b; \\
E(T_{J2}) & = b\sqrt{1 -  \rho^2}; \\
E(T_{MS}) = E(T_{JS}) &= (a+b)\sqrt{\frac{1+\rho}{2}}. 
\end{align} 
\end{subequations}
Figure \ref{fig:3Dsurface_ETs} gives the 3-D surfaces of the three SNRs at $b=1$, $a \in [-1, 1]$, and $\rho \in [-1, 1]$. 

Based on the magnitude and directionality of $a$, $b$, and $\rho$, which represent the signal and correlation patterns, we can conclude the relative advantages of these three test statistics based on their SNRs. In case of sparse signals, i.e., when $a=0$, $E(T_{M2})=b$ is always the largest. This result is consistent with Theorems \ref{thm.SNR.LM} -- \ref{thm.iHC.optimality.GLM} that marginal fitting, which is essentially the IT in linear models, is the optimal strategy for sparse signals.  

In case of dense signals, i.e., assuming $0 < |a| \leq b$, the relative advantage depends on $a$ and $\rho$. A few interesting observations can be made.  First, under independence with $\rho=0$, $E(T_{M2}) = E(T_{J2}) = b$ and $E(T_{MS}) = \frac{a+b}{\sqrt{2}}$. Here the marginal-fitting and the joint-fitting are equivalent; both could be better or worse than the summational statistic, depending on the values of $a$ and $b$. 
Secondly, when $a=b$, representing the equal-signal case (which is often assumed in theoretical studies, e.g., Theorem \ref{thm.iHC.optimality.GLM} above and literature studies such as \cite{Donoho2004, Hall2010}), the SNRs can be written as
\begin{align*}
E(T_{J2}) & = b\sqrt{1 -  \rho}\sqrt{1 +  \rho}; \\
E(T_{M2}) &= b(1+\rho) = b\sqrt{1 + \rho}\sqrt{1 +  \rho}; \\
E(T_{MS}) &= b\sqrt{2}\sqrt{1+\rho}. 
\end{align*} 
In this case, $E(T_{MS})$ is always the largest, indicating that statistic summation is the best, no matter what the correlation $\rho$ is. Meanwhile, if $\rho>0$, $E(T_{J2})$ is the smallest; if $\rho<0$, $E(T_{M2})$ is the smallest. 
Thirdly, consider positive dependence with $\rho > 0$, which is often true in many practical applications. In this case if $a>0$ has the same sign (representing the same signal direction) as $b$, then $E(T_{J2})$ is always the smallest, indicating that the joint-fitting is the least preferred in this case. 

It is also of interest to see the influence of correlation $\rho$ within each testing strategy. If SNR increases when comparing itself under $\rho=0$, then the signal is strengthened by correlation; otherwise, signal is weakened. First, for the joint model-fitting, note that $E(T_{J2}) =b\sqrt{1 -  \rho^2} < b$ no matter whether $\rho$ is positive or negative. This result indicates that correlation always weakens a signal under the joint model-fitting, which is irrelevant to the directionality of $\rho$ and $a$. 
Secondly, for the summational statistics, $E(T_{MS})$ in (\ref{equ.SNR.2D}) is monotone in $\rho$. That is, the signal is weakened when $\rho < 0$, and is strengthened when $\rho > 0$, which is true at any given $a$. Meanwhile, SNR is always weakened by cancellation whenever $a$ and $b$ have opposite signs at any given $\rho$. For example, if $a=-b$, $E(T_{MS})=0$, no matter whether $X_1$ and $X_2$ are positively or negatively correlated. In fact, unless $a$, $b$, and $\rho$ have the same sign,  $E(T_{MS})$ would always suffer signal cancellation. Moreover, if signals have different directions, $E(T_{MS})$ suffers more cancellation than $E(T_{M2})$. For example, when $a=-b$, we always have $E(T_{MS})=0$ indicating no signal, whereas $E(T_{M2})$ still is nonzero unless $\rho=1$. 
Thirdly, $E(T_{M2})$ depends on the product $a \rho$. If $a\rho$ is in the same direction as $b$, signal is enhanced; otherwise, signal is weakened by cancellation. 

Based on the comparisons among the three SNRs  in (\ref{equ.SNR.2D}), we can conclude a few general rules. First, the marginal fitting has the advantage when signals are sparser than the correlation, or when signals and data correlations are in the same direction (e.g., $a$, $b$, and $\rho$ have the same sign). Secondly, joint fitting could be preferred when marginal fitting leads to heavy signal cancellation. For example, when signals have the same sign but are negatively correlated, or when the signals have opposite signs but are positively correlated (e.g., $\rho=0.5, a=-1, b=1$ give $E(T_{J2}) =\sqrt{0.75} > E(T_{M2})= 0.5 > E(T_{MS})=0$). Thirdly, when $\rho \to -1$, $E(T_{J2}) = E(T_{MS}) = 0$ (i.e., signal fully cancelled), but marginal statistics still have a residual signal: $E(T_{M2})=b-a$, which could be even bigger if $a<0$. 

\begin{figure}[!t] 
	\begin{center}
		\subfloat{\includegraphics[width=2.1in,height=2.1in]{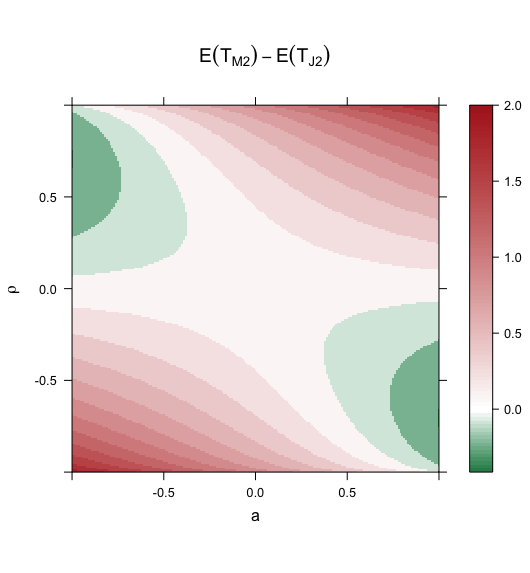}}
		\subfloat{\includegraphics[width=2.1in,height=2.1in]{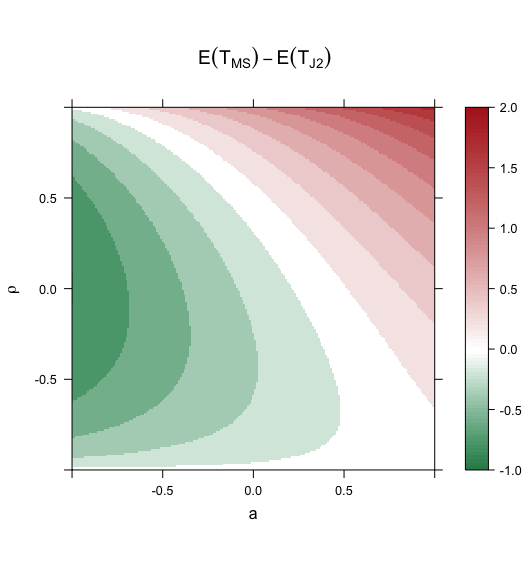}}
		\subfloat{\includegraphics[width=2.1in,height=2.1in]{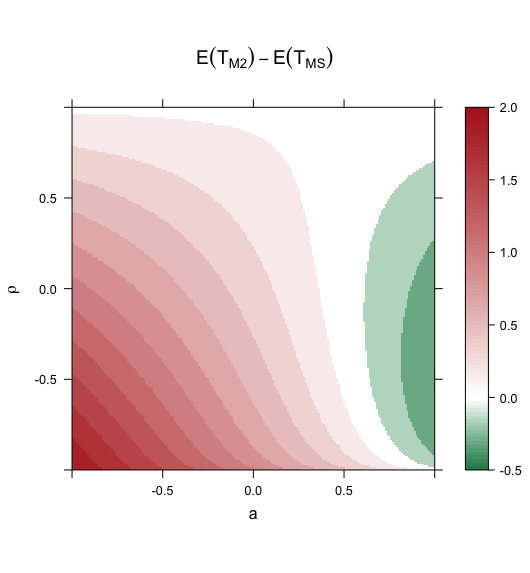}}
	\end{center}
	\caption{Contrasts of SNRs for marginal- and joint-fitting statistics and the summational statistic. 
	X-axis: $a \in [-1, 1]$; y-axis: $\rho \in [-1, 1]$; fixing $b=1$.}
\label{fig:contrast_ETs}
\end{figure}

Figure \ref{fig:contrast_ETs} shows the contrasts among these three SNRs in (\ref{equ.SNR.2D}) at $b=1$ with $a \in [-1, 1]$ and $\rho \in [-1, 1]$. The positive differences are red colored, indicating the first SNR is larger than the second one. The green color represents the opposite. For example, $E(T_{M2})$ is mostly equal or larger than $E(T_{J2})$, except in two relatively smaller areas, where $a$ and $\rho$ have opposite signs and both are large in magnitude. $E(T_{M2})$ is also mostly equal or larger than $E(T_{MS})$, except in one region where $a$ is positive and large. The joint-fitting and the summational statistic have roughly half-half advantages, for which $E(T_{MS})$ is larger than $E(T_{J2})$ mostly in the upper diagonal area with $a \geq -\rho$. 

\section{Numerical Studies}\label{Sect.numerics}

In this section we report the accuracy of $p$-value calculation methods, and then compare statistical power for the gGOF procedures under the GMM and the GLM with various correlation structures. 

\subsection{Accuracy of $p$-value Calculations\label{Sect.Psimu}}

\begin{figure}[!t] 
	\begin{center}
		\subfloat{\includegraphics[width=2.1in,height=2.1in]{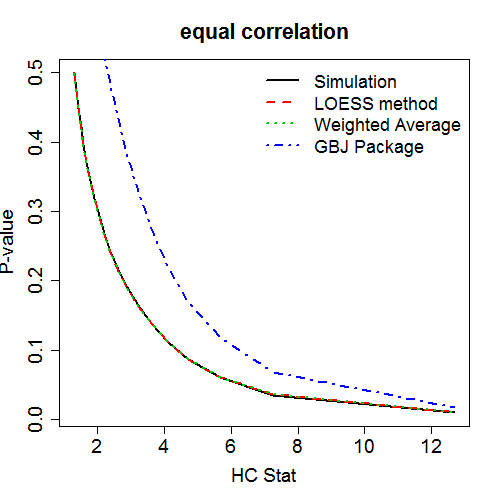}}
		\subfloat{\includegraphics[width=2.1in,height=2.1in]{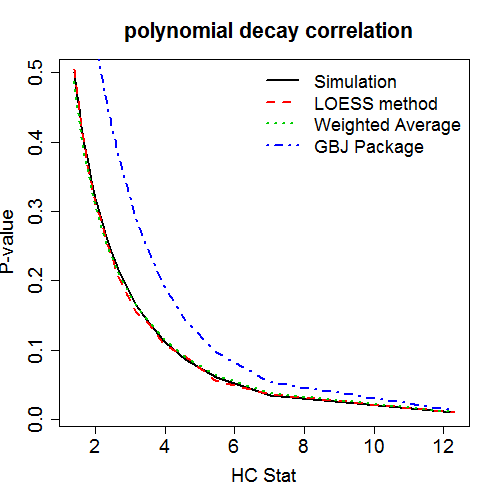}}
		\subfloat{\includegraphics[width=2.1in,height=2.1in]{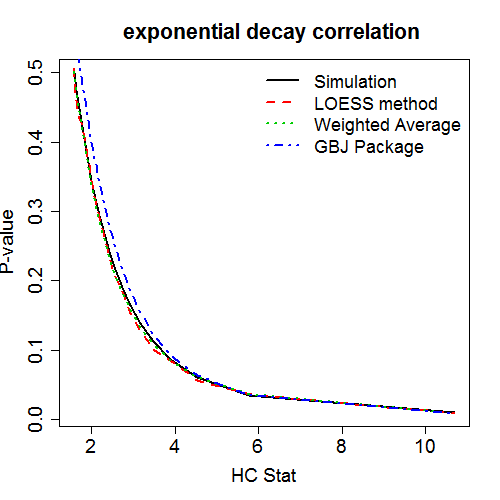}}\\
		\subfloat{\includegraphics[width=2.1in,height=2.1in]{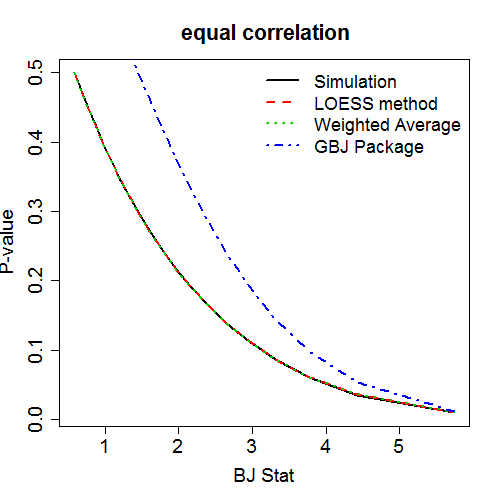}}
		\subfloat{\includegraphics[width=2.1in,height=2.1in]{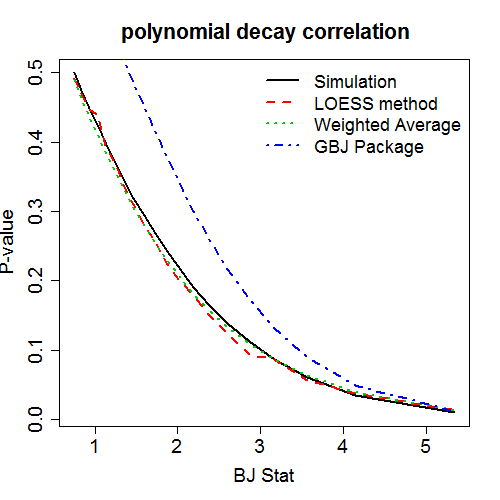}}
		\subfloat{\includegraphics[width=2.1in,height=2.1in]{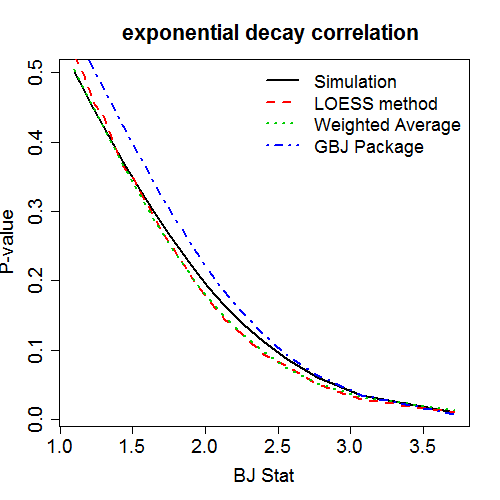}}
	\end{center}
	\caption{The survival function of the HC (first row) and the BJ (second row) under three correlation settings. Left: equal-correlation with $\rho=0.5$. Middle: polynomial-decay with $\gamma=0.5$. Right: exponential-decay with $\gamma=0.5$. $n=20$. $50,000$ repetitions for simulation.}
\label{fig:pval_GHC}
\end{figure}

First, we focused on the GMM in (\ref{equ.GMM}) with correlation matrix $\Sigma =  (\rho_{ij})_{1 \leq i, j \leq n}$. Three correlation structures were considered. The first is the equal-correlation model with $\rho_{ij}=\gamma$, for all $i \neq j$. The second is the polynomial-decay model with 
\begin{equation}
\label{equ.poly.decay}
\rho_{ij}=\frac{1}{(1+|i-j|)^\gamma}.
\end{equation}
The third is the exponential-decay model with $\rho_{ij}=\gamma^{|i-j|}$. According to how fast the correlation reduces, the equal-correlation model has the densest correlation and the exponential-decay model is the sparsest. We compared our $p$-value calculation methods with the moment-matching method \citep{barnett2016generalized, sun2017set}. The R package GBJ was applied for implementing the moment-matching method. 
We studied the HC (i.e., $HC^{2004}$) in (\ref{eq.HC.statistics}) with $\mathcal{R} =\{i:  1 \leq i \leq n \}$ and the BJ defined by (\ref{equ.phiDiverg.oneside}) with $s=1$ and $\mathcal{R} =\{i:  1 \leq i \leq n/2 \}$. Figure \ref{fig:pval_GHC} gives the survival functions under $H_0$, which can be considered as the curves of $p$-values over thresholds on the x-axis. It is clear that when the correlation is weaker, all methods are closer to the gold-standard obtained by simulation. However, our methods (both the WAM and the LOESS) are largely more accurate than the moment-matching method under stronger correlations.  

Next, we studied the multiple regression model in (\ref{equ.LM}). Under $H_0: \beta=0$, the input statistics have zero means but their correlation matrix depends on the data, as shown in Theorem \ref{thm.SNR.LM}. For simplicity, we first assumed no control covariates, i.e., $Z=0$, and the error variance $\sigma^2 = 1$. The design matrix $X_{N \times n}$ contained $N=1000$ observations and $n=20$ inquiry covariates grouped into two blocks: $A$ and $B$, each contained $10$ covariates. Seven correlation structures were considered and summarized in Table \ref{cor_types}. Covariates within and between blocks could have equal- or polynomial-decay correlations. Such block-wise correlation structures are a typical pattern in many real data applications. For example, genetic data often show haplotype blocks on DNA \citep{wall2003haplotype}. Cases 1--4 follow the same correlations as those studied in \cite{sun2017set}. To mimic the reality that subjects are often randomly sampled from population in different studies, we generated $X$ from the multivariate normal distribution in each of the $5,000$ simulations. Such simulation allows us evaluating the type I error rate control over meta studies. Based on the input statistics by the marginal model-fitting in (\ref{equ.Tmarg}), three gGOF tests were examined: the HC (i.e., $HC^{2004}$), the BJ, and an adaptive omnibus test adapting to these two statistics, all with $\mathcal{R}=\{1 \leq i \leq n/2\}$. Three typical significance levels at $\alpha = 0.05, 0.01$ and $0.005$ were considered. Table \ref{typeI_err} gives the empirical type I error rates, i.e., the proportion of simulated data sets that had the calculated $p$-values (by the LOESS method) being equal or less than $\alpha$.  The results show that our calculations were mostly accurate.  

\begin{table}[]
\centering
\caption{Seven correlation patterns among the covariates of  multiple regression model. Covariates have $2$ blocks: $A$ and $B$. $\rho=0.5$ for equal-correlation; decay: polynomial-decay correlation in (\ref{equ.poly.decay}) with $\gamma=1$. 
}
\label{cor_types}
\begin{tabular}{@{}llllllll@{}}
\toprule
 Cases         & 1 & 2      & 3      & 4      & 5     & 6     & 7     \\ \midrule
Within A  & 0 & $\rho$ & $\rho$ & $\rho$ & decay & decay & decay \\
Within B  & 0 & 0      & $\rho$ & $\rho$ & 0     & decay & decay \\
Cross A B & 0 & 0      & 0      & $\rho$ & 0     & 0     & decay \\ \bottomrule
\end{tabular}
\end{table}

\begin{table}[]
\centering
\caption{Empirical type I error rates at significance levels of $\alpha$ under regression model. Covariates follow normal distribution with 7 cases of correlations listed in Table \ref{cor_types}. Three statistics: the HC, the BJ, and an omnibus adapting to these two statistics. $\mathcal{R}=\{1 \leq i \leq n/2\}$.}
\label{typeI_err}
\begin{tabular}{@{}llllllll@{}}
\toprule
Case 1 &        &        &       & Case 2 &        &        &      \\ \midrule
$\alpha$ & BJ    &  HC     & Omni  & $\alpha$ &  BJ    &  HC     & Omni \\
0.05     & 0.0492 & 0.0528 & 0.0508      & 0.05     & 0.0514 & 0.0534  & 0.0530     \\
0.01     & 0.0100 & 0.0092 & 0.0096      & 0.01     & 0.0104 & 0.0106 & 0.0096     \\
0.005    & 0.0052 & 0.0046 & 0.0054      & 0.005    & 0.0056 & 0.0054 & 0.0052     \\
\toprule
Case 3 &        &        &       & Case 4 &        &        &      \\ \midrule
$\alpha$ &  BJ    &  HC     & Omni  & $\alpha$ &  BJ    &  HC     & Omni \\
0.05     & 0.0590 & 0.0508 &  0.0576     & 0.05     & 0.0500 & 0.0516 & 0.0510     \\
0.01     & 0.0092 & 0.0096 &   0.0092    & 0.01     & 0.0096 & 0.0116 &  0.0090    \\
0.005    & 0.0048 & 0.0046 &  0.0048     & 0.005    & 0.0054 & 0.0060 &  0.0056    \\
\toprule
Case 5 &        &        &       & Case 6 &        &        &      \\ \midrule
$\alpha$ &  BJ    &  HC     & Omni  & $\alpha$ &  BJ    &  HC     & Omni \\
0.05     & 0.0494 & 0.0526 &  0.0518     & 0.05     & 0.0570 & 0.0494 &  0.0524    \\
0.01     & 0.0108 & 0.0100 &   0.0098    & 0.01     & 0.0098  & 0.0084  &  0.0096    \\
0.005    & 0.0074 & 0.0050  &   0.0060   & 0.005    & 0.0046  & 0.0040 &  0.0044    \\
\toprule
Case 7 &        &        &       &          &        &        &      \\ \midrule
$\alpha$ &  BJ    &  HC     & Omni  &          &        &        &      \\
0.05     & 0.0572 & 0.0530 & 0.0528 &          &        &        &      \\
0.01     & 0.0126  & 0.0102 & 0.0114 &          &        &        &      \\
0.005    & 0.0050 & 0.0056 & 0.0040 &          &        &        &      \\ \bottomrule
\end{tabular}
\end{table}

\begin{figure}[!t] 
	\begin{center}
		\subfloat{\includegraphics[width=3in]{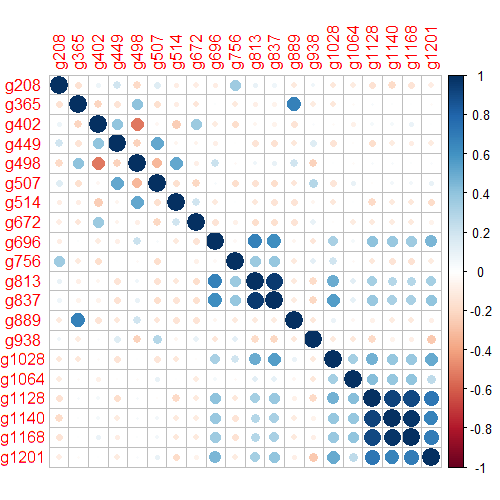}}
	\end{center}
	\caption{Correlation matrix for genotypes of the 20 simulated common variants by Cosi2.}
\label{fig:pval_GWAS_cov}
\end{figure}

Furthermore, we evaluated the $p$-value calculation methods by simulations in the context of genetic studies. Specifically, we obtained 1,290 haplotypes for a genome region of 250k base-pairs by a genetical data simulation software Cosi2 \citep{shlyakhter2014cosi2}. The genetical model follows the typical coalescent model with the linkage disequilibrium (LD) pattern being similar to those of the European population. 
Two haplotypes were randomly drawn with replacement from the 1,290 haplotypes to form each subject's genotypes. The sample sizes considered were $N=200$, $500$ or $1000$. The first $n=20$ common variants, with minor allele frequency (MAF) $>$ 5\%, were used as the inquiry genetic covariates. An example of the correlation structure among 20 genotypes is illustrated in Figure \ref{fig:pval_GWAS_cov}.  
Two non-genetic control covariates were further simulated: $Z_1$ is a binary variable of Bernoulli$(0.5)$ and $Z_2$ is a continuous variable of $N(0, 1)$. The responses were generated from the non-genetic control covariates in order to mimic the null hypothesis that the phenotype is influenced only by environmental factors, not by any genetic variants:
\begin{align*}
Y = 0.5 Z_1 + 0.1 Z_2 + \epsilon,
\end{align*}
where $\epsilon \sim N(0, 1)$. We examined five gGOF tests: the HC, the BJ, the reverse BJ, the reverse HC (i.e., $HC^{2008}$), and a digGOF test adapting to both these four statistics and six truncation domains $\mathcal{R}_j=\{k_{0j} \leq i \leq k_{1j}\}$ with $k_{0j}\in\{1,2\}$ and $k_{1j}\in\{10,15,20\}$.  Table \ref{typeI_err_gwas} summarizes the empirical type I errors rate, i.e., the proportion of simulations where the calculated $p$-values (by the LOESS method based on simulated test statistics) were less than the given $\alpha$ value. The results show that calculations were mostly accurate, or sometimes slightly conservative because the true type I error rates, represented by the empirical values, were slightly smaller than the nominal $\alpha$ values.

\begin{table}[]
\centering
\caption{Empirical type I error rates at significance levels of $\alpha$ under simulated genetic data. 
}
\label{typeI_err_gwas}
\begin{tabular}{@{}lllllllllll@{}}
\toprule
 &         &        &       $N=200$&  &        &        \\ \midrule
$\alpha$ & BJ    &  HC     &   rev. BJ& rev. HC & digGOF    \\
0.05     & 0.0662 & 0.0522 & 0.0471      & 0.0421     & 0.0573  \\
0.01     & 0.0091 & 0.0102 & 0.0049      & 0.0058     & 0.0111 \\
0.005   & 0.0038 & 0.0050 & 0.0017     & 0.0023    & 0.0060  \\
0.001   & 0.0003 & 0.0009 & 0.0002      & 0.0003    & 0.0009  \\\midrule
 &        &        &       $N=500$&  &        &        \\ \midrule
$\alpha$ & BJ    &  HC     &   rev. BJ& rev. HC & digGOF    \\
0.05     & 0.0665 & 0.0545 & 0.0433      & 0.0442     & 0.0605  \\
0.01     & 0.0089 & 0.0116 & 0.0050      & 0.0051     & 0.0129 \\
0.005   & 0.0045 & 0.0051 & 0.0015      & 0.0014   & 0.0065 \\
0.001   & 0.0002 & 0.0013 & 0.0003      & 0.0004   & 0.0015 \\\midrule
 &         &        &       $N=1000$&  &        &        \\ \midrule
$\alpha$ &  BJ    &  HC     &   rev. BJ& rev. HC & digGOF    \\
0.05     & 0.0675 & 0.0540 & 0.0539      & 0.0466     & 0.0614 \\
0.01     & 0.0094 & 0.0100 & 0.0055      & 0.0079     & 0.0130 \\
0.005   & 0.0042 & 0.0048 & 0.0013     & 0.0024   & 0.0076 \\
0.001   & 0.0006 & 0.0009 & 0.0003      & 0.0003   & 0.0014 \\
\bottomrule
\end{tabular}
\end{table} 


\subsection{Comparisons of Statistical Power}\label{Sect.CompPower}

\begin{figure}[!t] 
	\begin{center}
		\subfloat{\includegraphics[width=2.5in,height=2.5in]{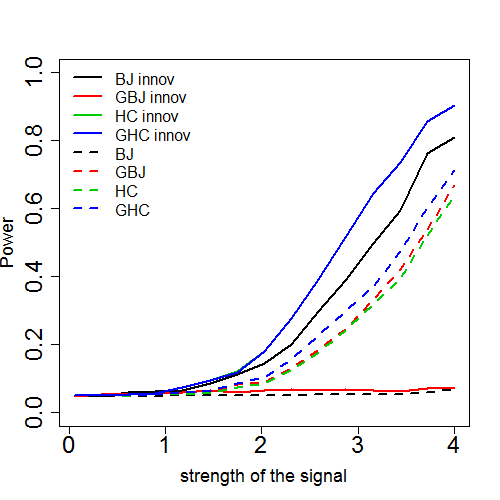}}
		\subfloat{\includegraphics[width=2.5in,height=2.5in]{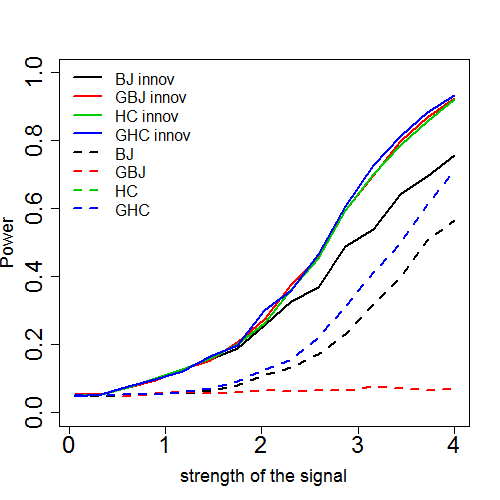}}\\
		\subfloat{\includegraphics[width=2.5in,height=2.5in]{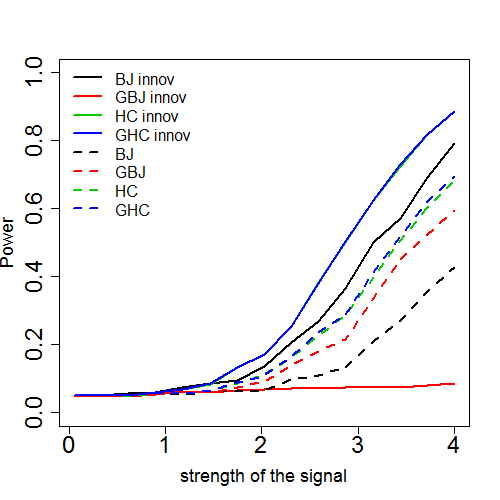}}
		\subfloat{\includegraphics[width=2.5in,height=2.5in]{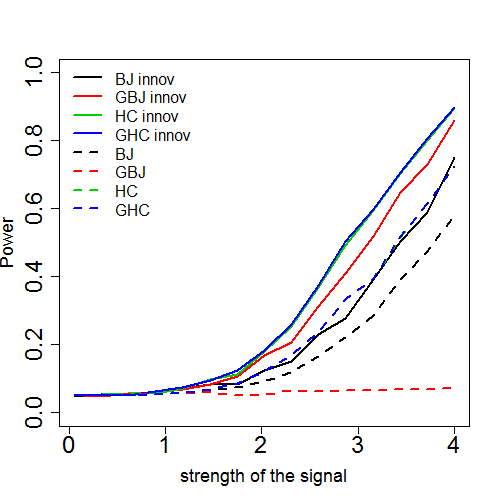}}
	\end{center}
	\caption{Power comparison among the HC, the GHC, the BJ and the GBJ for detecting the single signal. X-axis: strength of the signal.  Top-left penal: $\Sigma^{(1)}_{ij}=0.3$ for all $i\neq j$; top-right: $\Sigma^{(2)}_{ij}=-0.01$ for all $i\neq j$; bottom-left: $\Sigma^{(3)}_{ij}=|i-j|^{-1}$ for all $i\neq j$; bottom-right: $\Sigma^{(4)}= (\Sigma^{(3)})^{-1}$ (i.e., polynomial-decay with negative entries). For all: $n=100$.}
\label{fig:power_GM_sig_strength}
\end{figure}

To study statistical power under the GMM, the input statistics $T$ were generated from the multivariate normal distribution with various mean vector $\mu$ and correlation matrices $\Sigma$. First, we considered the single-signal case, i.e.,  only one nonzero element exists in $\mu$. Figure \ref{fig:power_GM_sig_strength} gives power curves for  the HC, the GHC, the BJ and the GBJ tests, where the input statistics were either the original $T$ ($n=100$; SNR values over x-axis) or the IT-transformed $T$.  Four correlation structures were studied: positive equal-correlation (top-left penal: $\Sigma^{(1)}_{ij}=0.3$); negative equal-correlation (top-right: $\Sigma^{(2)}_{ij}= -0.01$);  
positive polynomial-decay (bottom-left: $\Sigma^{(3)}_{ij}=|i-j|^{-1}$); and negative polynomial-decay (bottom-right: $\Sigma^{(4)}= (\Sigma^{(3)})^{-1}$). Recall that the analytical results in Section \ref{Sect.IT.sparse} indicate that for the single-signal case the IT always increases SNR. Here, the simulation illustrates that the IT indeed always increases the power of the HC, the GHC and the BJ. The HC and the GHC give almost identical power after the IT (their curves almost overlapped), which is  the highest in the four settings. However, the performance of the GBJ depends on the correlation structure. For example, when correlations are positive equal-correlation or polynomial-decay (shown in left panels), the GBJ almost have no power after the IT. Note that in these two settings the IT-transformed $T$ has negative correlations. 

Next, we considered increasing the number of signals with their locations being randomly distributed at each simulation. Figure \ref{fig:power_GM_sig_number} gives power curves under the same four correlation structures as those in Figure \ref{fig:power_GM_sig_strength}. The IT always increases statistical power  of the HC, the GHC and the BJ at any numbers of signals. However, in some cases (e.g., in the two left panels) under sparse signals, the GBJ's power could be even reduced after the IT. 
Furthermore, after the IT, the BJ is always similar as or significantly better than the GBJ, and the HC is always similar as or slightly better than the GHC.  	

Now we move to power study under the GLM setting. The genotype data were simulated with the design matrix $X_{N\times n}$, where sample size $N=1000$ and the number of SNPs $n=100$. All SNPs had MAF = 0.3, i.e., $X_{ij} \sim$binomial$(2, 0.3)$. Consider two covariate blocks: block A contained all trait-associated SNPs (i.e., true signals) with block-size $n_A \in \{1, 2, ..., 10\} $; block B contained all non-associated SNPs with block size $n_B=n-n_A$.  Denote equal-correlations within block A by $\rho^A_{ij}$, those within block B by $\rho^B_{ij}$, and those cross A and B by $\rho^{AB}_{ij}$. Four correlation structures were defined: 
\begin{enumerate}
\item[(A)] $\rho^A_{ij} = 0.5, \rho^B_{ij}=0, \rho^{AB}_{ij}=0$; 
\item[(B)] $\rho^A_{ij} = 0.5, \rho^B_{ij}=0.2, \rho^{AB}_{ij}=0$; 
\item[(C)] $\rho^A_{ij} = 0.5, \rho^B_{ij}=0.2, \rho^{AB}_{ij}=0.2$; 
\item[(D)] $\rho^A_{ij} = 0.5, \rho^B_{ij}=0.2, \rho^{AB}_{ij}=0.33$.
\end{enumerate}
The effect of the trait-associated SNPs have equal effect size $\beta^A_j$ that is chosen to give comparable powers. See Appendix Table \ref{effect_sizes} for details. Assuming there were no control covariates, we considered quantitative traits, where the response variable $Y$ was obtained by regression model in (\ref{equ.LM}) with $\sigma=1$. We also considered binary traits, where $Y$ was obtained by the GLM model in (\ref{equ.the GLM}) with the logistic link function. We ran 5,000 simulations, and the significance level was set at 0.05. 
 
\begin{figure}[!t] 
	\begin{center}
		\subfloat{\includegraphics[width=2.5in,height=2.5in]{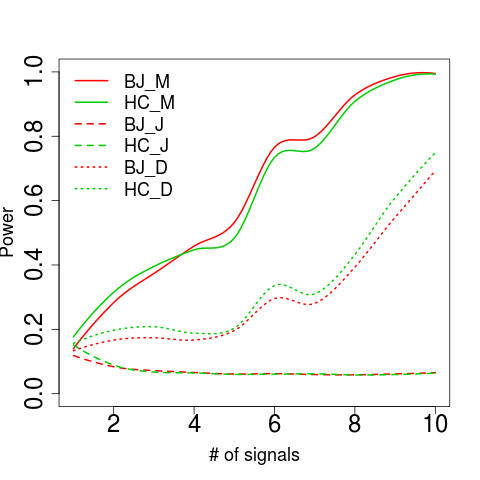}}
		\subfloat{\includegraphics[width=2.5in,height=2.5in]{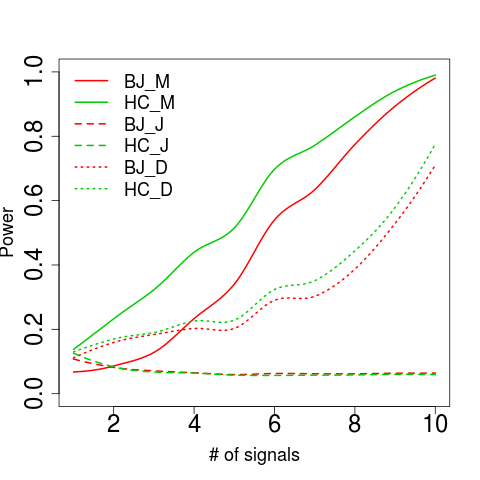}}\\
		\subfloat{\includegraphics[width=2.5in,height=2.5in]{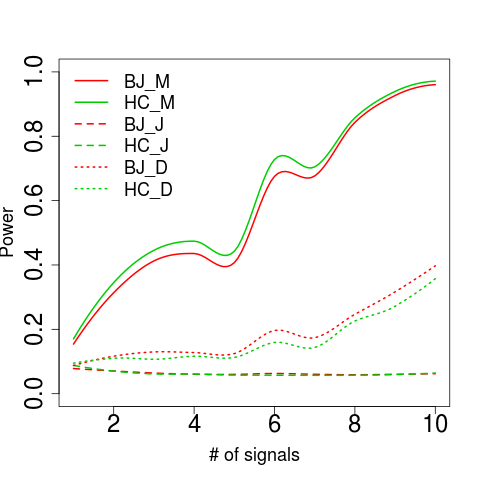}}
		\subfloat{\includegraphics[width=2.5in,height=2.5in]{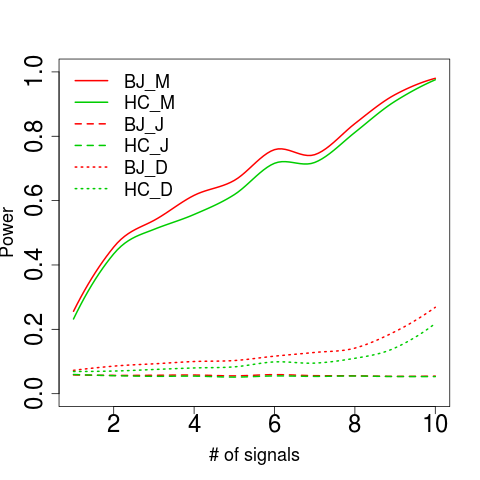}}
	\end{center}
	\caption{Statistical power of the HC and the BJ under different model-fitting of linear regression with genetic covariates. J: the joint model-fitting; M: the marginal model-fitting (i.e., the IT); D: de-correlation. Four correlation structures: (A): top-left panel; (B): top-right; (C): lower-left; (D): lower-right. 	}
\label{fig:newpower_sigprop_equal_margvsjoint}
\end{figure}

First, we studied how statistical power would be affected by types of model-fittings. Considering the HC and the BJ for quantitative trait, under these positive correlations Figure \ref{fig:newpower_sigprop_equal_margvsjoint} shows that overall the joint model-fitting has the lowest power, the marginal model-fitting (i.e., the IT) has the highest power, and the de-correlation transformation of either gives the intermediate power. We also studied the performance of the GHC and the GBJ as illustrated in Appendix Figure \ref{fig:newpower_sigprop_equal_margvsjoint_GBJ}, which shows the similar relative performance. Note that if the data correlations were negative though, it is still possible that such clustered positive signals would less favor the marginal model-fitting due to signal cancellations, as illustrated in the left panel of Figure \ref{fig:contrast_ETs}. 

\begin{figure}[!t] 
	\begin{center}
		\subfloat{\includegraphics[width=2.5in,height=2.5in]{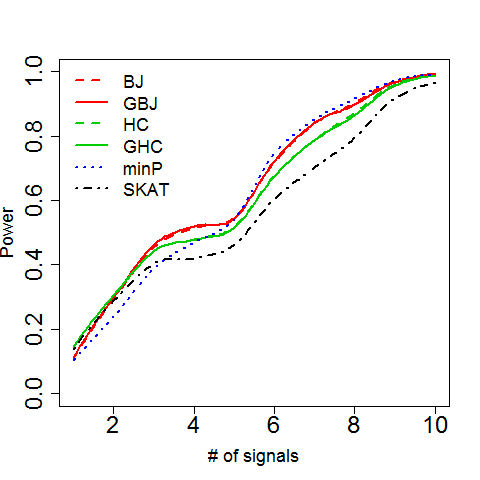}}
		\subfloat{\includegraphics[width=2.5in,height=2.5in]{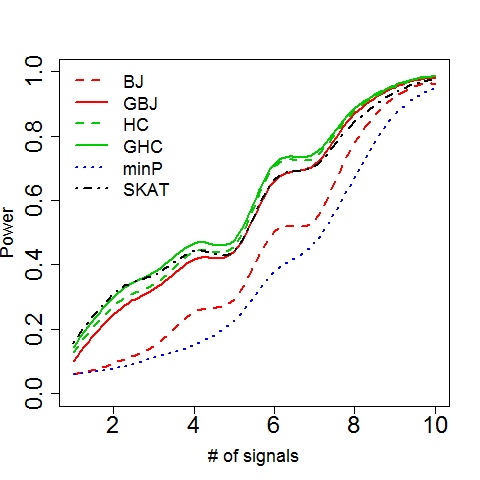}}\\
		\subfloat{\includegraphics[width=2.5in,height=2.5in]{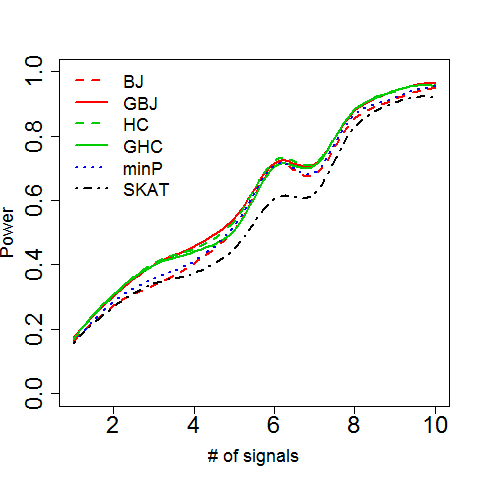}}
		\subfloat{\includegraphics[width=2.5in,height=2.5in]{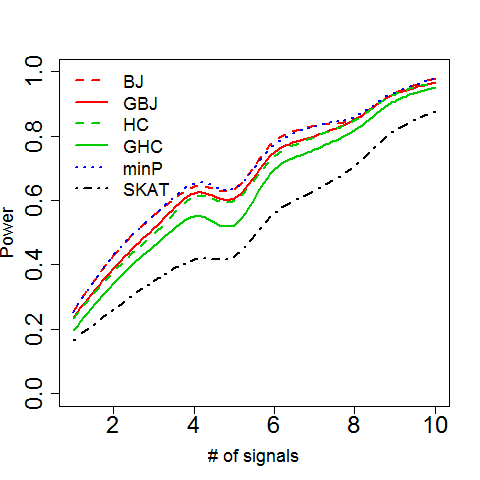}}
	\end{center}
	\caption{Power of statistics under linear regression of genetic covariates. Four correlation structures: (A): top-left panel; (B): top-right; (C): lower-left; (D): lower-right.}
\label{fig:newpower_sigprop_equal}
\end{figure}

\begin{figure}[!t] 
	\begin{center}
		\subfloat{\includegraphics[width=2.5in,height=2.5in]{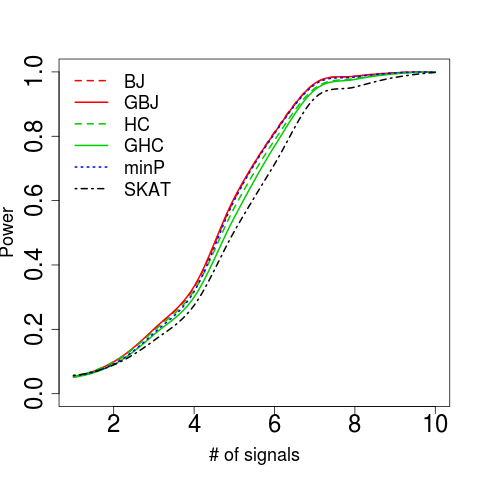}}
		\subfloat{\includegraphics[width=2.5in,height=2.5in]{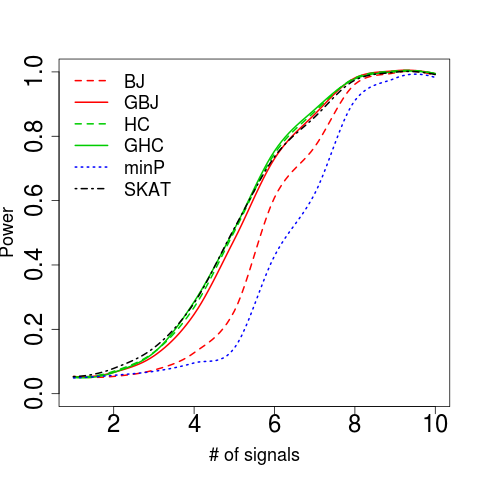}}\\
		\subfloat{\includegraphics[width=2.5in,height=2.5in]{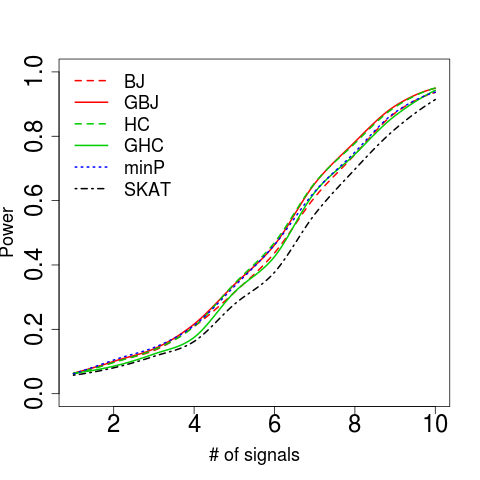}}
		\subfloat{\includegraphics[width=2.5in,height=2.5in]{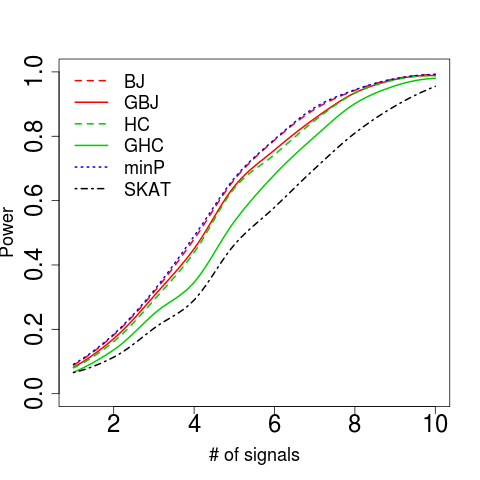}}
	\end{center}
	\caption{Power of statistics under logistic regression of genetic covariates.  Four correlation structures: (A): top-left panel; (B): top-right; (C): lower-left; (D): lower-right.}
\label{fig:newpower_sigprop_equal_the GLM}
\end{figure}

Based on the marginal model-fitting, Figures \ref{fig:newpower_sigprop_equal} and \ref{fig:newpower_sigprop_equal_the GLM} show power comparisons under regression and logit models, respectively. We compared six methods: the HC, the GHC, the BJ, the GBJ, the minimal $p$-value method, and the SKAT \citep{wu2011rarevariant}. A few observations can be made. First, in the correlation case (A) where only causal SNPs are correlated, the HC's performance overlaps with the GHC, and the BJ overlaps with the GBJ. Here the HC/GHC is better for sparser signals (i.e., fewer causal SNPs) and the BJ/GBJ are better for denser signals. This observation is consistent with above results and the literature \citep{sun2017set, Zhang2016distributions}. Secondly, in the correlation case (B) where the signals and noises have within correlations but no between-correlations, the GBJ is more powerful than the BJ. However when the correlations between blocks A and B increase, as is shown in the correlation cases (C) and (D), the HC and the BJ outperform the GHC and the GBJ, respectively. 

\begin{figure}[!t] 
	\begin{center}
		\subfloat{\includegraphics[width=2.5in,height=2.5in]{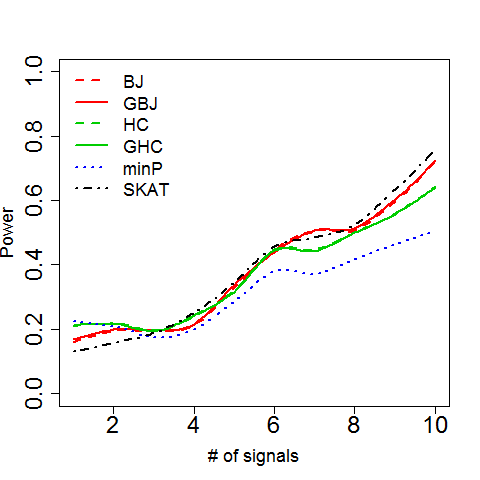}}
		\subfloat{\includegraphics[width=2.5in,height=2.5in]{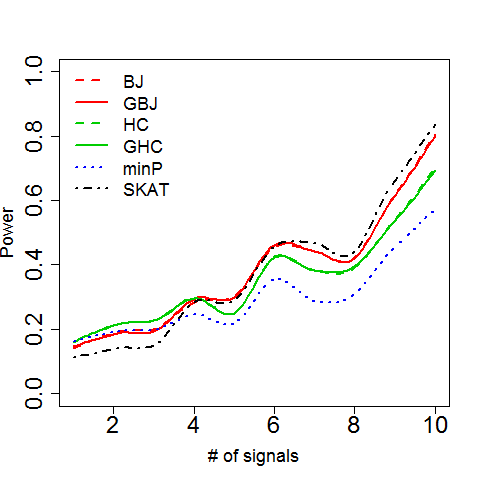}}\\
		\subfloat{\includegraphics[width=2.5in,height=2.5in]{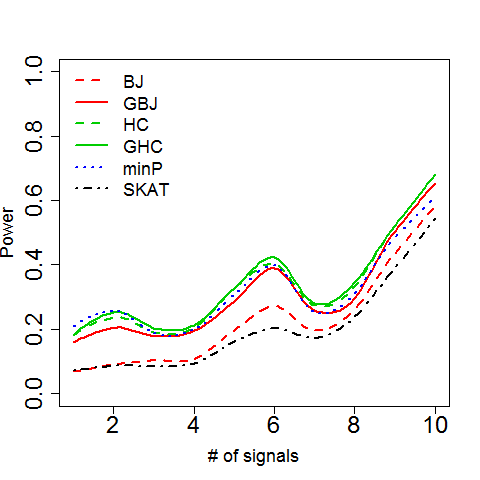}}
		\subfloat{\includegraphics[width=2.5in,height=2.5in]{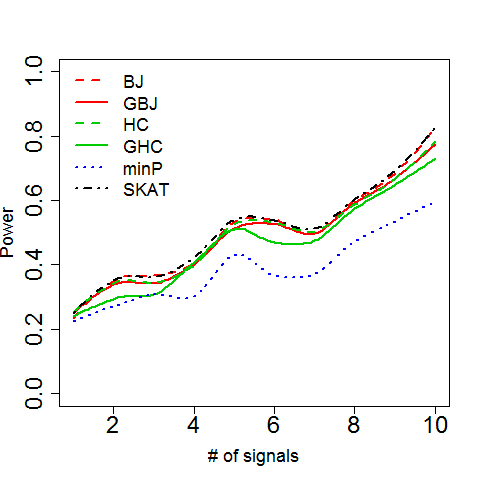}}
	\end{center}
	\caption{Power comparison under regression model with genotype covariates and block-wise equal-correlations. Correlation structures follow Cases 1 -- 4 in Table \ref{cor_types} with $\rho=0.3$.  Case 1: top-left panel, i.e., independent covariates; Case 2: top-right; Case 3: lower-left; Case 4: lower-right.}
\label{fig:power_sigprop_equal}
\end{figure}

\begin{figure}[!t] 
	\begin{center}
		\subfloat{\includegraphics[width=2.1in,height=2.1in]{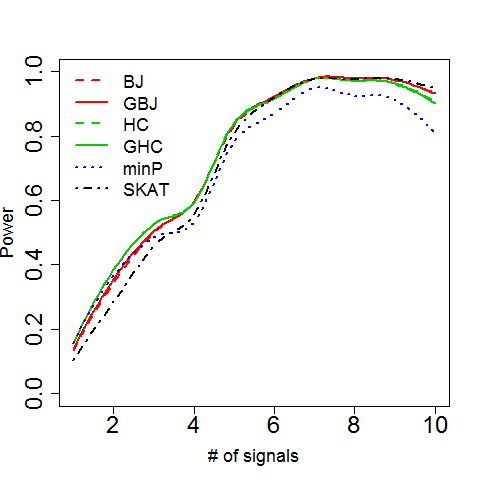}}
		\subfloat{\includegraphics[width=2.1in,height=2.1in]{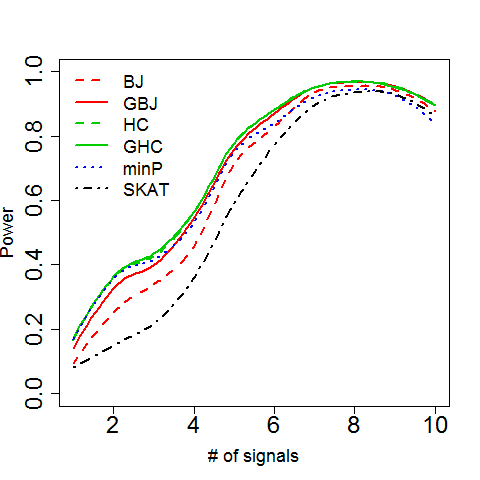}}
		\subfloat{\includegraphics[width=2.1in,height=2.1in]{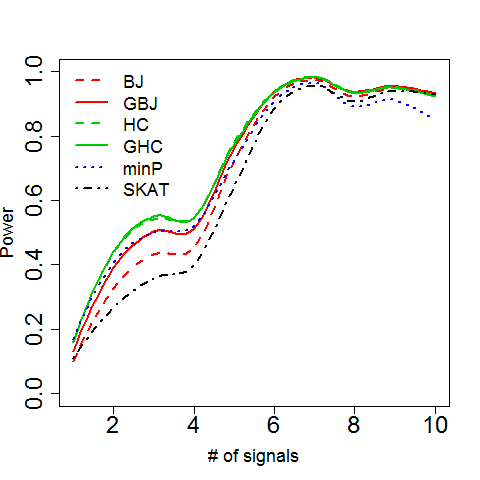}}
	\end{center}
	\caption{Power comparison under regression model with genotype covariates and polynomial-decay correlations. Left, middle, and right panels follow correlation structures Cases 5 -- 7 in Table \ref{cor_types}, respectively, with $\gamma=1$. 
	}
\label{fig:power_sigprop_poly}
\end{figure}

We further studied the seven correlation patterns in Table \ref{cor_types}. For cases 1--4 of equal-correlations, Figure \ref{fig:power_sigprop_equal} shows that the results support the same conclusion that when the between-correlations are large, the HC and the BJ could be more powerful than the GHC and the GBJ, respectively.  For cases 5 --7 of polynomial-decay correlations, Figure \ref{fig:power_sigprop_poly} shows that the HC and the GHC are virtually equivalent, and give highest power over most signal proportions.  

Besides the genotype covariates, we also studied continuous covariates under the GLM. To be consistent, we followed the same settings for Figures \ref{fig:power_sigprop_equal} and \ref{fig:power_sigprop_poly}, except each covariate in $X$ followed $N(0,1)$. Appendix Figures \ref{fig:power_linreg_equal} and \ref{fig:power_linreg_poly} correspond to equal-correlations and polynomial-decay correlations, respectively. Similar patterns were observed as those in Figures \ref{fig:power_sigprop_equal}  and \ref{fig:power_sigprop_poly}. The similarity evidences that the distribution of the design matrix $X$ (either discrete of Gaussian) is not a primary influential factor for the performance.
More influential factors are the SNR and correlation structure.

\section{Application to Genetic Association Studies}\label{Sect.GWAS}

In this section we exam the $p$-value calculation for the gGOF statistics in gene-based SNP-set association studies. SNPs are often in LD and thus their input statistics are correlated. Two real data sets were studied. The first study is a GWAS for Crohn's disease. The data contains 1,145 individuals from non-Jewish population (572 Crohn's disease cases and 573 controls) \citep{Duerr2006}. After typical quality control for genotype data, 308,330 somatic SNPs were grouped into 15,857 genes according to their physical locations. The logit model was applied to get the input statistics for the gGOF. The control covariates $Z$ contained the intercept and first two principal components of the genotype data in order to control potential population structure \citep{Price2006}. 
The second study is a whole exome sequencing study for the ALS. The data-cleaning and SNP-filtering processes followed the original study \citep{smith2014exome}. 
The final data contains 457 ALS cases and 141 controls, with 105,764 SNPs in 17,088 genes. Two non-genetic categorical covariates, gender and country origin (6 countries), were included as the control covariates $Z$ in the SNP-association tests based on the logit model. For both data analyses no gGOF test was needed for genes containing only one SNP. 

\begin{figure}[h] 
	\begin{center}
		\includegraphics[width=2.2in,height=2.2in]{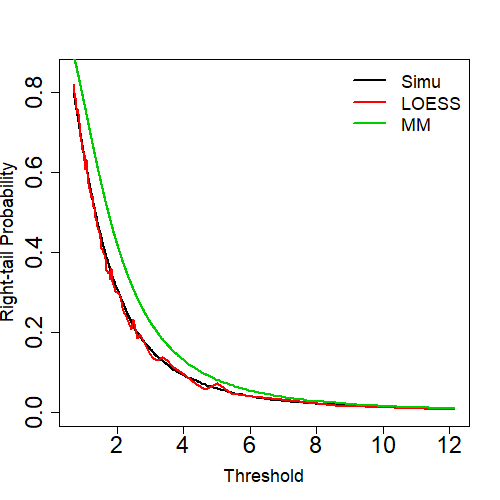}
		\includegraphics[width=2.2in,height=2.2in]{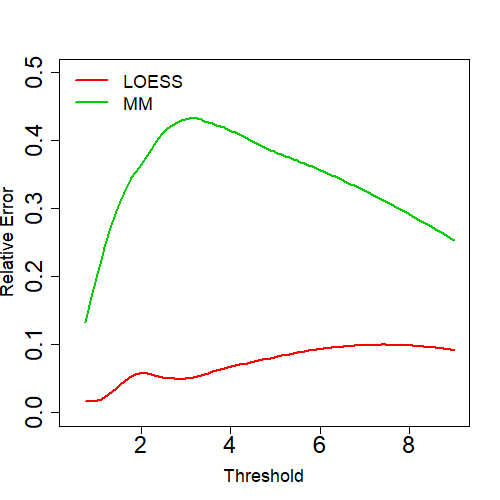}\\
		\includegraphics[width=2.2in,height=2.2in]{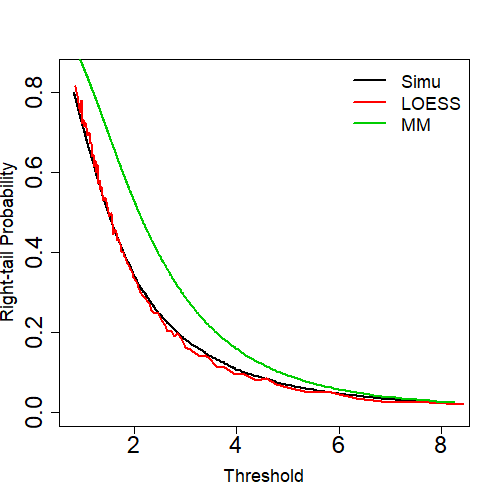}
		\includegraphics[width=2.2in,height=2.2in]{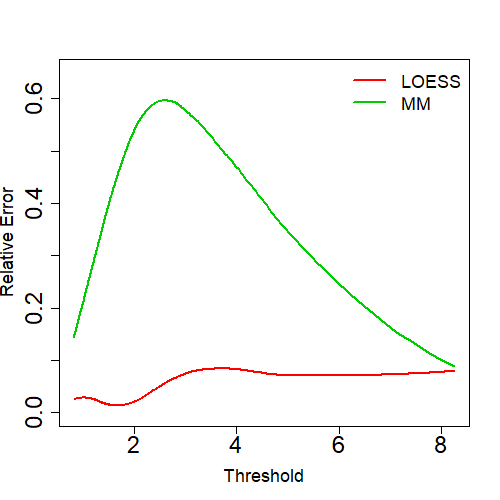}\\
		\includegraphics[width=2.2in,height=2.2in]{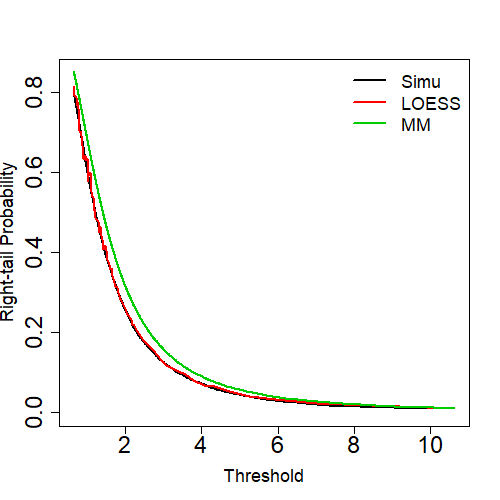}
		\includegraphics[width=2.2in,height=2.2in]{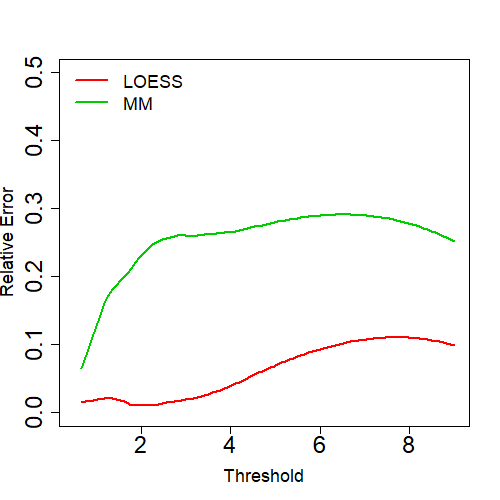}
	\end{center}
\caption{Calculation accuracy for the null distribution of the HC by the real genotype data of Crohn's disease study. Simu: by simulation, the gold standard; MM: the moment-matching method; LOESS: the LOESS method. Row 1: {\it CARD15}; Row 2: {\it IL23R}; Row 3: {\it FLJ20558}. Left: Null distribution curves; Right: Relative errors to simulation.}  
\label{fig:geneexample}
\end{figure}

First, by comparing with simulations we assessed the $p$-value calculation accuracy based on real genotype data with typical gene sizes and LD structures.  Figure \ref{fig:geneexample} shows examples of three genes  {\it CARD15}, {\it IL23R}, and {\it FLJ20558} based on the GWAS data of Crohn's disease. Comparing with the moment-matching based method (by the R package GBJ), our calculation method is generally better. 

\begin{figure}[!t] 
	\begin{center}
		\subfloat{\includegraphics[width=2.5in,height=2.5in]{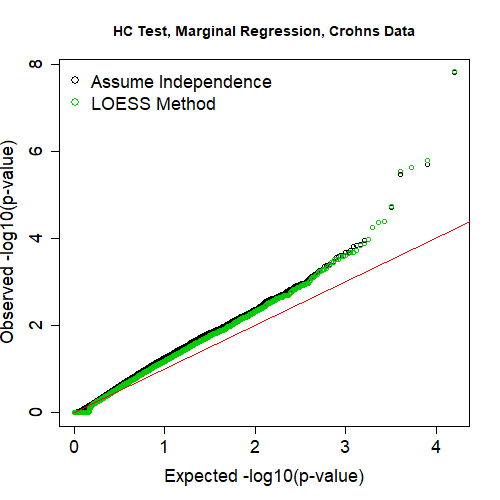}}
		\subfloat{\includegraphics[width=2.5in,height=2.5in]{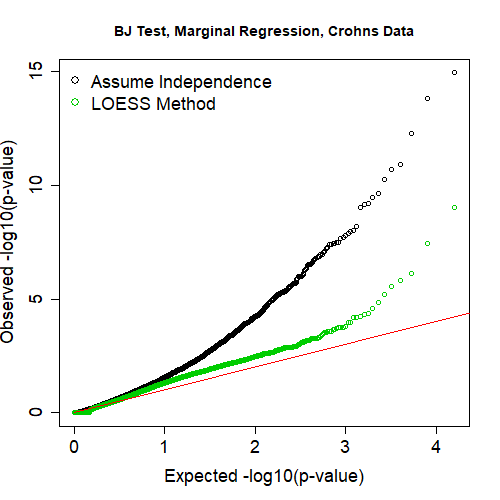}}\\
		\subfloat{\includegraphics[width=2.5in,height=2.5in]{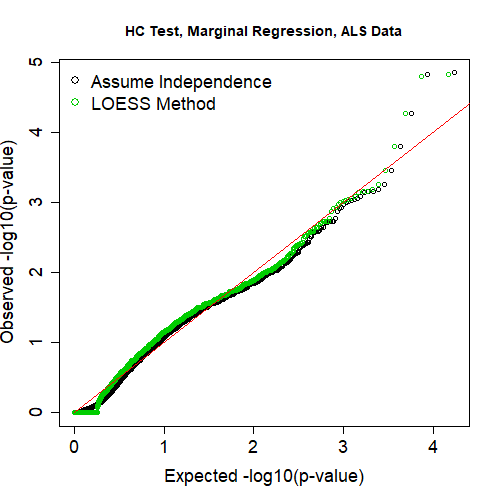}}
		\subfloat{\includegraphics[width=2.5in,height=2.5in]{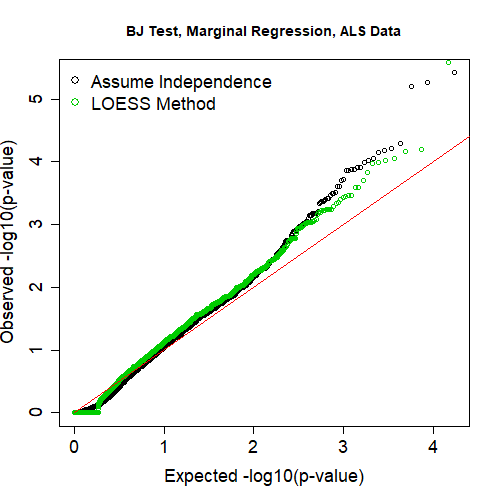}}
	\end{center}
	\caption{
	QQ plots of the HC and the BJ statistics. P-value calculations were based on independence assumption vs. correlation-incorporated calculations by LOESS.  Top: Crohn's disease data. Bottom: ALS disease data. Left: the HC test. Right: the BJ test.}
\label{fig:gwas}
\end{figure}

Secondly, for all genes we calculated the $p$-values when incorporating SNP correlations estimated by their genotype data. The results were compared with the calculation that ignores data correlation, i.e., by the method under independence \citep{Zhang2016distributions}. Figure \ref{fig:gwas} shows that the correlation-incorporated calculation successfully brought the QQ plots closer to the diagonals, indicating a better type I error rate control than the correlation-ignored calculation. Meanwhile, comparing with the HC, the BJ is more sensitive to correlations, so the correlation-incorporated calculation is more important to the BJ. In general, sequencing data have less correlations than GWAS data, a fact that made the ALS QQ plots being closer to the diagonal. However, the correlation-incorporated calculation still showed benefit, especially for the BJ. Furthermore, some top ranked genes are known to be relevant disease genes. For example, {\it CARD15} is a well-known gene for Crohn's disease \citep{hugot2001association}, and {\it SMAP1} and {\it KIAA1377} are related to the ALS \citep{andres2017amyotrophic, keogh2013next}. Other putative genes need further study to confirm biological connections. 

\section{Discussion}\label{Sect.Discuss}

This paper provides a unified solution for a generic family of goodness-of-fit statistics in analysis of correlated data. Comparing to individual developments, the study of a broad family possesses a few advantages. First, it allows immediate applications of any existing or newly developed gGOF type statistics to analyzing correlated data. Secondly, different statistics could have advantages over different situations. Therefore, as a group the family retains high statistical power over a broad parameter space. For example, the HC and the GHC often have similar high power for sparser signals. 
Other methods such as the reverse HC and the reverse BJ could be preferred when signals are weaker and/or denser. Thirdly, the double-adaptation gGOF, which adapts to both the proper statistic function and the proper $p$-value truncation scheme for a given data, can be robust over different situations. Even more interestingly, the distribution of the omnibus test is shown to be a cross-boundary probability. Thus analytical $p$-value calculation can be unified without the typical need of computationally intensive simulations.  

One challenge of studying correlated data is that it relays on the joint distribution of the input statistics.  This paper illustrated how correlation structure and signal patterns work together to influence SNR and thus the power of signal-detection methods. In particular, not only signal magnitudes, but also signal locations in relation to the dependence structure are relevant. We set a starting point for defining the IT: the signal locations should be independent of the correlations before transformation. Under this condition, this paper showed that the marginal model-fitting is the IT of the joint model-fitting under the GLM. Since the IT is shown often preferred for analyzing correlated data, even though the marginal model-fitting gives biased estimation for the coefficients, it often gives higher power for the signal-detection purpose. Meanwhile, through a comprehensive study of SNR formulas under the bivariate regression model, we demonstrated relative advantages of four fundamental modeling strategies: marginal-fitting, joint-fitting, de-correlation, and statistic summation. The results provide an insightful understanding, for example, on how dense signals and correlations could lead to relative advantages or drawbacks of these strategies. We proposed the digGOF methodology, which is novel and potentially significant to broad data analysis applications. 

Beyond the achievements of this paper, there are a few future studies we plan to carry on.  First, this paper assumed that correlation matrix can be properly estimated. In real data analysis, this requirement could be a challenge, for example, in high dimensional data where the number of covariates is much larger than the sample size. Such problem may be easier to handle when correlations are relatively sparse \citep{bickel2008regularized, cai2010optimal}, but otherwise it could be much more difficult. Secondly, this paper studied the case that the input statistics are Gaussian or close to Gaussian. We showed that the assumption is reasonable under the GLM when sample size is large. It is of interest to study cases when the input statistics are non-Gaussian too.  

\acks{We would like to acknowledge partial support for this project from the National Science Foundation (NSF grants DMS-1309960 and DMS-1812082). }



\appendix \label{Sect.appendix}
\section*{Appendix A: Proofs.}

This appendix contains proofs of the main theorems. 

\subsection*{Proof of Theorem \ref{thm.H0distn.equalCorr}}

Consider input statistics $T$ in (\ref{equ.GMM}) with $\mu=1$ and $\Sigma_{ij} = \rho$. The elements of $T$ can be written as 
\[
T_i = \rho Z_i + \sqrt{1- \rho^2}Z_0, \quad i = 1, ..., n,
\] 
where $Z_0, Z_1, ..., Z_n \overset{iid}{\sim} N(0, 1)$. For simplicity, we consider $\mathcal{R} = \{i:  k_0 \leq i \leq k_1 \}$. Let $U_{(1)} \leq ... \leq U_{(n)}$ be the order statistics of $n$ {\it iid} Uniform$(0,1)$ random variables. First we consider that input $p$-values in (\ref{equ.GMM.pValue}) are one-sided. For a the gGOF statistic $S_{n, f, \mathcal{R}}$  in (\ref{equ.gGOFstat}), its distribution function can be written in (\ref{equ.cross.bound.prob}) -- (\ref{equ.rejectionBoundary}). By the property of exchangeable normal random variables \citep{tong2012multivariate} we can get the cumulative distribution function (CDF):  
\begin{align*}
\mathbb{P}(S_{n, f, \mathcal{R}} < b)&=\mathbb{P}(P_{(i)}>u_i \text{, all } i=k_0,...,k_1) \\\
&=\mathbb{P}(T_{(n-i+1)}<\Phi^{-1}(1-u_i), i=k_0,...,k_1)\\\
&=\mathbb{P}(\sqrt{1-\rho}Z_{(n-i+1)}+\sqrt{\rho}Z_0<\Phi^{-1}(1-u_i), i=k_0,...,k_1)\\\
&=\int_{-\infty}^\infty \phi(z)\mathbb{P}(Z_{(n-i+1)}<\frac{\Phi^{-1}(1-u_i)-\sqrt{\rho}z}{\sqrt{1-\rho}}, i=k_0,...,k_1)dz\\\
&=\int_{-\infty}^\infty \phi(z)\mathbb{P}(U_{(i)}>1-\Phi\left(\frac{\Phi^{-1}(1-u_i)-\sqrt{\rho}z}{\sqrt{1-\rho}}\right),i=k_0,...,k_1)dz.
\end{align*}
When the input $p$-values are two-sided, the calculation is adjusted accordingly. 
\begin{align*}
\mathbb{P}(S_{n, f, \mathcal{R}} < b)&=\mathbb{P}(P_{(i)}>u_i, i=k_0,...,k_1) \\\
&=\mathbb{P}(2(1-\Phi(|T|_{(n-i+1)}))>u_i, i =k_0,...,k_1)\\
&=\mathbb{P}(|\sqrt{1-\rho}Z+\sqrt{\rho}Z_0|_{(n-i+1)}<\Phi^{-1}(1-u_i/2),i=k_0,...,k_1)\\\
&=\int_{-\infty}^\infty \phi(z_0)\mathbb{P}(|\sqrt{1-\rho}Z+\sqrt{\rho}z_0|_{(n-i+1)}<\Phi^{-1}(1-u_i/2), i=k_0,...,k_1)dz_0\\\
&=\int_{-\infty}^\infty \phi(z_0)\mathbb{P}(U_{(i)}>1-F_{z_0}\left(\Phi^{-1}(1-u_i/2)\right), i=k_0,...,k_1)dz_0, 
\end{align*}
where $F_{z_0}$ is the CDF of $|\sqrt{1-\rho}Z+\sqrt{\rho}z_0|$: 
\[
F_{z_0}\left(x\right) = \Phi\left(\frac{x-\sqrt{\rho}z_0}{\sqrt{1-\rho}}\right)-\Phi\left(\frac{-x-\sqrt{\rho}z_0}{\sqrt{1-\rho}}\right). 
\]
In summary, define $c_{1i}=1-\Phi\left(\frac{\Phi^{-1}(1-u_i)-\sqrt{\rho}z}{\sqrt{1-\rho}}\right)$, $c_{2i}=1-F_{z}\left(\Phi^{-1}(1-u_i/2)\right)$, then 
\[
\mathbb{P}(S_{n, f, \mathcal{R}} < b)=\int_{-\infty}^\infty \phi(z)\mathbb{P}(U_{(i)}>c_{t i}, i=k_0,...,k_1)dz, \quad{}t=1,2. 
\]

\subsection*{Proof of Theorem \ref{thm.SNR.LM}}

Now we prove Theorem \ref{thm.SNR.LM}. First we derive the least squares estimator of $\beta$ by the joint model-fitting. Write $S_{N\times(n+m)}=[X,Z]$ and $\alpha_{(n+m)\times 1}=[\beta^\prime, \gamma^\prime]^\prime$,
\begin{align*}
	\hat{\alpha} =  (S^\prime S)^{-1}S^\prime  Y \Longleftrightarrow
 \begin{bmatrix}
		\hat{\beta}\\
		\hat{\gamma}
     \end{bmatrix} = 
 \begin{bmatrix}
		X^\prime  X &X^\prime  Z\\
		Z^\prime  X &Z^\prime  Z
     \end{bmatrix} ^{-1}
 \begin{bmatrix}
		X^\prime  Y\\
		Z^\prime  Y
     \end{bmatrix} .
\end{align*}
We can write the matrix inverse explicitly,
\begin{align*}
	&\begin{bmatrix}
		X^\prime  X &X^\prime  G\\
		G^\prime  X &G^\prime  G
     \end{bmatrix} ^{-1}\\
	=
	&\begin{bmatrix}
		(X^\prime  (I-H)X)^{-1}&-(X^\prime  (I-H)X)^{-1}X^\prime  Z(Z^\prime  Z)^{-1}\\
		-(Z^\prime  Z)^{-1}Z^\prime  X(X^\prime  (I-H)X)^{-1} &(X^\prime  X)^{-1}-(X^\prime  X)^{-1}Z^\prime  X(X^\prime  (I-H)X)^{-1}X^\prime  Z(X^\prime  X)^{-1}
     \end{bmatrix} 
\end{align*}
Thus 
\begin{align*}
	\hat{\beta}_J &= 
	(X^\prime  (I-H)X)^{-1}X^\prime  Y - (X^\prime  (I-H)X)^{-1}X^\prime  Z(Z^\prime  Z)^{-1}Z^\prime  Y\\
	&=(X^\prime  (I-H)X)^{-1}X^\prime  (I-H)Y
\end{align*}
Since the convariance matrix of $Y$ is $I$ and $I-H$ is idempotent, the covariance matrix of $\hat{\beta}$ is 
\begin{align*}
	{\rm Var}(\hat{\beta}_J) &= (X^\prime  (I-H)X)^{-1}X^\prime  (I-H)\sigma^2 I (I-H)X(X^\prime  (I-H)X)^{-1}\\
	&= \sigma^2 (X^\prime  (I-H)X)^{-1}
\end{align*}
Thus the scaled test statistics by the joint model-fitting is $T_J$ given in (\ref{equ.Tjoint.reg}) with mean $\mu_{T_J} = \Lambda\beta$ and correlation matrix $\Sigma_{T_J} = \Lambda(X^\prime(I-H)X)^{-1}\Lambda$. 

For the marginal model-fitting between $Y$ and the $j$th covariate $X_j$, $j=1, ..., n$: 
\[
\hat{Y}_{Mj} =X_j \hat{\beta}_{Mj} + Z \hat{\gamma},  
\]
following the same deduction as above, the least squares estimator is 
\[
\hat{\beta}_{Mj} = (X'_j(I-H)X_j)^{-1}X'_j(I-H)Y \sim N(\frac{1}{X'_j(I-H)X_j}X'_j(I-H)X\beta, \frac{\sigma^2}{X'_j(I-H)X_j}).
\]
The corresponding scaled statistic is 
\[
T_{Mj} = \frac{1}{\sigma}\sqrt{X'_j(I-H)X_j} \hat{\beta}_{Mj} = \frac{1}{\sigma\sqrt{X'_j(I-H)X_j}}X'_j(I-H)Y \sim N(\frac{1}{\sigma\sqrt{X'_j(I-H)X_j}}X'_j(I-H)X\beta, 1),
\]
and the vector $T_M = (T_{M1}, ..., T_{Mn})$ is
\[
T_M = CX'(I-H)Y/\sigma \sim N(CX'(I-H)X\beta/\sigma,  CX'(I-H)XC) = N(\Sigma_{T_M} C^{-1}\beta/\sigma, \Sigma_{T_M}).
\] 

Next, we consider the transformations. To simplify the deduction, define $X^* = (I-H)X\Lambda^{-1}$. We can write 
\begin{align*}
T_J &= \Lambda(X^\prime(I-H)X)^{-1}X^\prime(I-H)Y = (X^{*\prime}X^*)^{-1}X^{*\prime}Y /\sigma; \\
\Sigma_{T_J} &= \Lambda(X^\prime(I-H)X)^{-1}\Lambda = (X^{*\prime}X^*)^{-1};\\
D_{T_J} &= {\rm diag}(\frac{1}{\sqrt{(\Sigma_{T_J}^{-1})_{jj}}})_{1 \leq j \leq n} \text{ where } (\Sigma_{T_J}^{-1})_{jj} = \Lambda^{-1}_j X_j^\prime(I-H)X_j \Lambda^{-1}_j.
\end{align*}
Apply the IT defined in (\ref{equ.IT.stat}) to $T_J$, 
\begin{align*}
T^{IT}_J = D_{T_J}\Sigma^{-1}_{T_J} T_J = D_{T_J} X^{*\prime}Y /\sigma,
\end{align*}
for which the $j$th element is 
\[
T^{IT}_{Jj} = \frac{1}{\sqrt{(\Sigma_{T_J}^{-1})_{jj}}} X^{*\prime}_j Y /\sigma = \frac{1}{\sqrt{\Lambda^{-1}_j X_j^\prime(I-H)X_j \Lambda^{-1}_j}} \Lambda^{-1}_j X_j^\prime(I-H)Y/\sigma = T_{Mj}. 
\]

Regarding the DTs, for the marginal fitting
\[
T_M^{DT} = U_M T_M = U_M C X'(I-H)Y /\sigma. 
\]
and since
\[
I = U_M \Sigma_{T_M} U'_M = U_M C X'(I-H)X C U'_M.
\]
The mean is $E(T_M^{DT}) = U_M C X^\prime(I-H) X\beta/\sigma = (CU'_M)^{-1} \beta/\sigma$. 
For the joint fitting, let $U_J$ be the inverse of the Cholesky factorization of $\Sigma_{T_J}$, i.e., 
\[
I = U_J \Sigma_{T_J} U'_J = U_J \Lambda (X'(I-H)X)^{-1}\Lambda U'_J.
\]
We have 
\[
(\Lambda U'_J)^{-1} = (C U'_M)' = U_M C. 
\]
Thus,
\[
T_J^{DT} = U_J T_J = U_J \Lambda  (X^\prime(I-H)X)^{-1} X^\prime(I-H)Y = (\Lambda U'_J)^{-1} X^\prime(I-H)Y = U_M C X^\prime(I-H)Y. 
\]

Lastly, for the SNRs, assume $\beta_{j}/\sigma =A >0$ and $\beta_i = 0$ for all $i \neq j$. Since $(U_J)_{j j} \geq 1$, we have 
\[
E(T_{Jj}^{DT}) = (U_J \Lambda \beta/\sigma)_{j} =  (U_J)_{j j} \Lambda_{jj} A \geq \Lambda_{jj} A = (\Lambda \beta/\sigma)_{j} = E(T_{Jj}). 
\]
Note also 
\[
E(T_{Mj}) = (CX'(I-H)X\beta/\sigma)_{j} = (\Sigma_{T_M} C^{-1}\beta/\sigma)_{j} = C^{-1}_{j j} A,
\]
and since $(U_M^{-1})_{j j} \leq 1$, we have
\[
E(T_{Jj}^{DT}) = ((U_M^{-1})'C^{-1}\beta/\sigma)_{j} \leq C^{-1}_{j j} A. 
\]
For all $i \neq j$, $E(T_{Ji})  = E(T_{Ji}^{DT})  = E(T_{Mi}) = 0$. This ends the proof. 

\subsection*{Proof of Theorem \ref{thm.SNR.the GLM}}

First, by the maximum likelihood estimation (MLE) and the Fisher Scoring method, we show that the MLE estimator of $\beta$ by the joint model-fitting is given in (\ref{eq.jointBeta.the GLM}). Define $S_{N\times(n+m)}=[X,Z]$ and $\alpha_{(n+m)\times 1}=[\beta^\prime, \gamma^\prime]^\prime$. The log-likelihood:
\begin{align*}
\log L = \sum_{i=1}^N \log L_i = \sum_{i=1}^N \left[\frac{y_i\theta_i-b(\theta_i)}{a_i(\phi)}+c(y_i,\phi)\right] 
\end{align*}
Thus the score functions,
\begin{align*}
U_j(\alpha) = \frac{\partial\log L}{\partial \alpha_j} = \sum_i \frac{\partial\log L_i}{\partial \theta_i}\frac{\partial\theta_i}{\partial \alpha_j}
\end{align*}
Since 
\begin{align*}
\frac{\partial\log L_i}{\partial \theta_i}=\frac{y_i-b^\prime(\theta_i)}{a_i(\phi)}=\frac{y_i-\mu_i}{a_i(\phi)}
\end{align*}
and with canonical link $\theta_i=g(\mu_i)=S_{i\cdot}\alpha$,
\begin{align*}
\frac{\partial\theta_i}{\partial \alpha_j}=\frac{\partial S_{i\cdot}\alpha}{\partial \alpha_j}= S_{ij}
\end{align*}
We have 
\begin{align}
U_j(\alpha) = \sum_{i=1}^N \frac{S_{ij}(Y_i-\mu_i)}{a_i(\phi)}, j=1,...,n+m
\label{equ.score}
\end{align}
or in vector form $U(\alpha) = A^{-1}S^\prime(Y-\mu)$, where $A={\rm diag}(a_i(\phi))$ which is a diagonal matrix of the overdispersion parameters. For linear regression and logistic regression $A$ is identity. 

Next we can show that the Fisher's information matrix
\begin{align}
\begin{split}
\mathcal{I}_n(j,k) = cov(U_j(\alpha), U_k(\alpha))
=-E\left[\frac{\partial U_j(\alpha)}{\partial\alpha_k}\right] 
= \sum_{i=1}^N S_{i\cdot}^\prime\frac{b^{\prime\prime}(\theta_i)}{a_i(\phi)}S_{i\cdot}
\end{split}
\label{equ.fisherinfo}
\end{align}
or in matrix form $\mathcal{I}_n=S^\prime W S$, where $W={\rm diag}(var(Y_i)/a_i^2(\phi))$. When there is no overdispersion, the weight matrix becomes a diagonal matrix of the variances of $Y_i$. 

The MLE estimator of $\alpha$ is obtained by solving the score equations $U(\alpha)=0$ through the Fisher Scoring method. 
\begin{align}
\alpha^{(j+1)} = \alpha^{(j)} + \left(\mathcal{I}_n^{(j)}\right)^{-1}U(\alpha^{(j)})
\label{equ.fisherscore}
\end{align}

Consider a one-step MLE (cf. Theorem 4.19 and Exercise 4.152 of \cite{shaostat}) with initial estimation
\begin{align*}
{\alpha^{(0)}}^\prime = (0, {\gamma^{(0)}}^\prime)
\end{align*}
where $\gamma^{(0)}$ is the MLE of $\gamma$ using only control covariates $Z$. 

For simplicity we focus on the case $A=I$ (no overdispersion), if not, we can re-define the $S$ to be $A^{-1}S$ and the arguments are the same.
\begin{align*}
\begin{pmatrix}
    \beta^{(1)}        \\
    \gamma^{(1)}       
\end{pmatrix}
=
\begin{pmatrix}
    0        \\
    \gamma^{(0)}       
\end{pmatrix}
+
\begin{pmatrix}
    X^\prime W^{(0)} X  &  X^\prime W^{(0)} Z    \\
    Z^\prime W^{(0)} X &   Z^\prime W^{(0)} Z   
\end{pmatrix}^{-1}
\begin{pmatrix}
    X^\prime(Y-\mu^{(0)})        \\
    Z^\prime(Y-\mu^{(0)})       
\end{pmatrix}
\end{align*} 
where $\mu^{(0)}$ is the MLE estimator of $\mu$ using only the control covariates $Z$. $W^{(0)}$ is the corresponding MLE estimator of the weight matrix defined in/below \ref{equ.fisherinfo}. 

Define $\tilde{X}={W^{(0)}}^{1/2}X$, $\tilde{Z}={W^{(0)}}^{1/2}Z$ and $\tilde{H}=\tilde{Z}(\tilde{Z}^\prime\tilde{Z})^{-1}\tilde{Z}^\prime$
\begin{align*}
\beta^{(1)} = \left(\tilde{X}^\prime(I-\tilde{H})\tilde{X}\right)^{-1}X^\prime(Y-\mu^{(0)}) - \left(\tilde{X}^\prime(I-\tilde{H})\tilde{X}\right)^{-1}\tilde{X}^\prime\tilde{Z}(\tilde{Z}^\prime\tilde{Z})^{-1}Z^\prime(Y-\mu^{(0)})
\end{align*}
The second term is exactly $0$ because $\mu^{(0)}$ is the MLE using only $Z$, thus the scores $Z^\prime(Y-\mu^{(0)})\equiv 0$. The MLE of $\beta$ therefore can be written as 
\begin{align}
\hat{\beta}_J = \left(\tilde{X}^\prime(I-\tilde{H})\tilde{X}\right)^{-1}X^\prime(Y-\mu^{(0)}) \overset{D}{\to} N(\beta, (\tilde{X}^\prime(I-\tilde{H})\tilde{X})^{-1})
\label{equ.betamle}
\end{align}
After standardization of $\hat{\beta}_J$, the test statistics by the joint model-fitting is $T_J$ in \ref{equ.Tjoint.the GLM}. The rest of the proof follows  the proof of Theorem~\ref{thm.SNR.LM}.

\subsection*{Proof of Theorem \ref{thm.iHC.optimality.GLM}}

Note that for all $1 \leq j \leq n$:
\[
(\Sigma^{-1})_{jj} = [( \tilde{\Lambda}( \tilde{X}^\prime(I- \tilde{H}) \tilde{X})^{-1} \tilde{\Lambda})^{-1}]_{jj} = \tilde{\Lambda}^{-2}_{jj} \tilde{X}_j^\prime(I- \tilde{H}) \tilde{X}_j = C(f) + o(1),
\]
where 
\[
C(f) = \frac{1}{2\pi} \int_{-\pi}^\pi f(t)^{-1} dt. 
\]
So, after the IT, the SNR is 
\[
\sqrt{C(f)}\tilde{\Lambda}_{jj}\beta_j \approx \sqrt{\tilde{X}_j^\prime(I- \tilde{H}) \tilde{X}_j} \beta_j = \sqrt{2\tilde{X}_j^\prime(I- \tilde{H}) \tilde{X}_j r_j \log n} > \sqrt{2 \rho(\alpha) \log n}.
\]
The result applies by following the definition of iHC in (4.8) and the assumptions and the conclusion of Theorem 5.1 in \cite{Hall2010}. 

\newpage
\section*{Appendix B: More Numerical Study Results}

This appendix contains extended numerical studies. 

\begin{figure}[H] 
	\begin{center}
		\subfloat{\includegraphics[width=2.5in,height=2.5in]{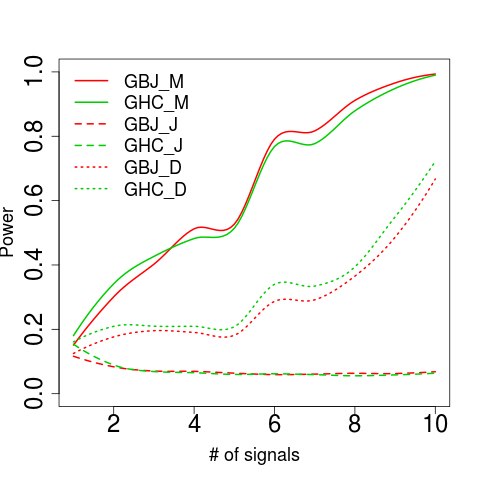}}
		\subfloat{\includegraphics[width=2.5in,height=2.5in]{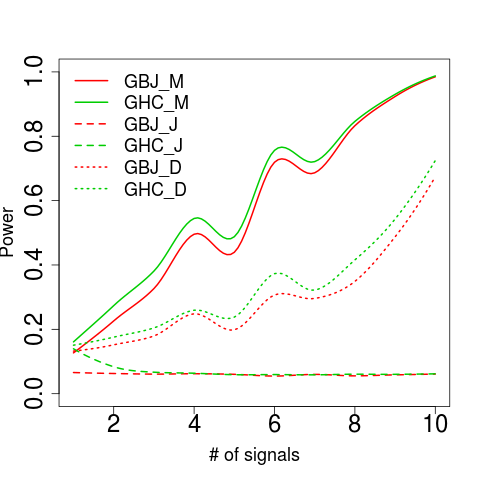}}\\
		\subfloat{\includegraphics[width=2.5in,height=2.5in]{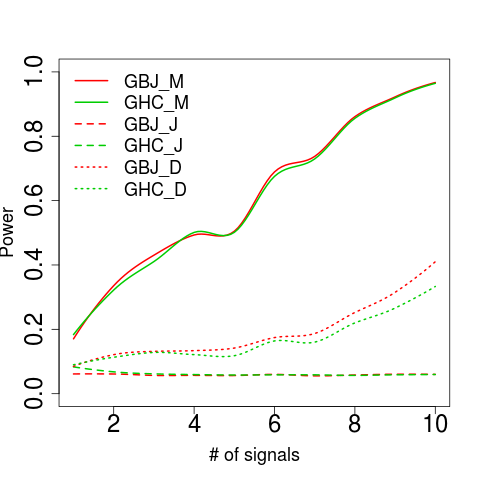}}
		\subfloat{\includegraphics[width=2.5in,height=2.5in]{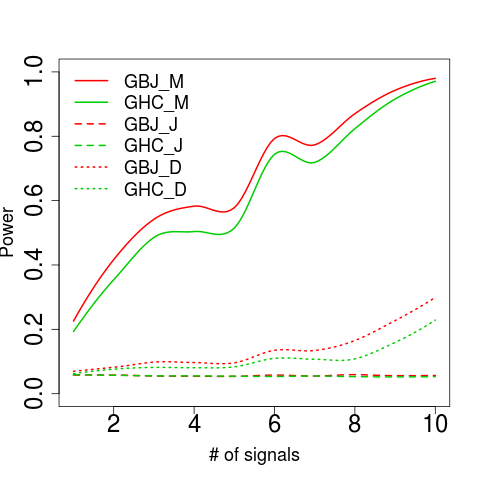}}
	\end{center}
	\caption{Statistical power of the GHC and the GBJ under different model-fitting of linear regression with genetic covariates. J: the joint model-fitting. M: the marginal model-fitting. D: de-correlation. Four correlation structures: (A): top-left panel; (B): top-right; (C): lower-left; (D): lower right.}
\label{fig:newpower_sigprop_equal_margvsjoint_GBJ}
\end{figure}

\begin{figure}[H] 
	\begin{center}
		\subfloat{\includegraphics[width=2.5in,height=2.5in]{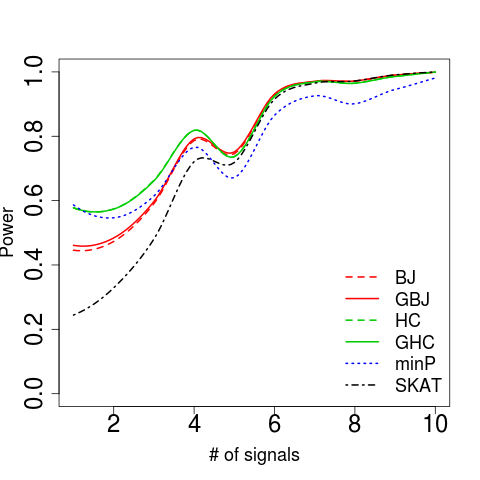}}
		\subfloat{\includegraphics[width=2.5in,height=2.5in]{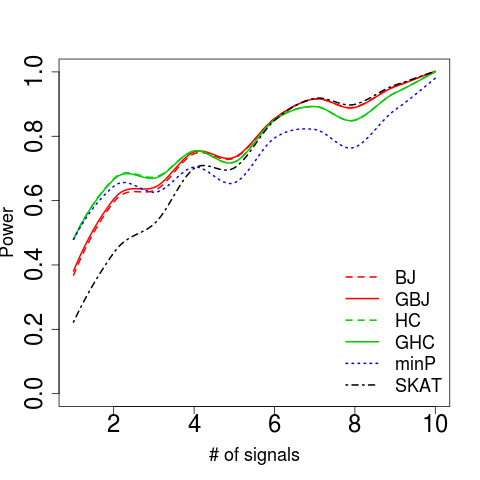}}\\
		\subfloat{\includegraphics[width=2.5in,height=2.5in]{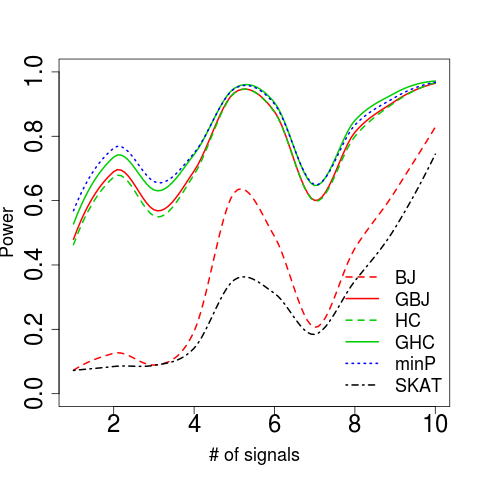}}
		\subfloat{\includegraphics[width=2.5in,height=2.5in]{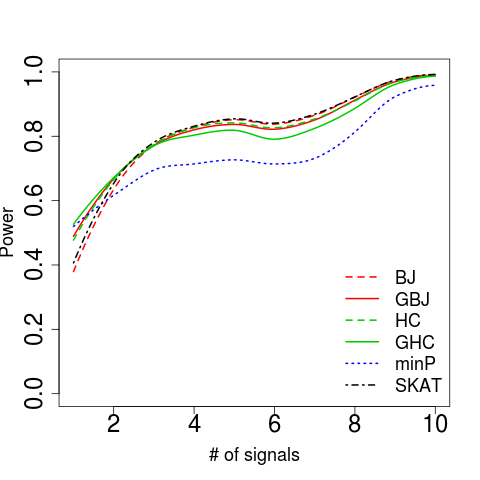}}
	\end{center}
	\caption{Power comparison under regression model with $N(0,1)$ distributed covariates and block-wise equal-correlations. Correlation structures follow Cases 1--4 in Table \ref{cor_types} with $\rho=0.3$. Case 1: top-left panel, i.e., independent covariates; Case 2: top-right; Case 3: lower-left; Case 4: lower-right.}
\label{fig:power_linreg_equal}
\end{figure}

\begin{figure}[H] 
	\begin{center}
		\subfloat{\includegraphics[width=2.1in,height=2.2in]{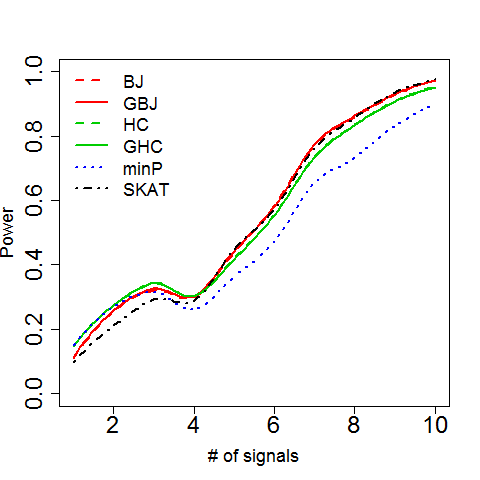}}
		\subfloat{\includegraphics[width=2.1in,height=2.2in]{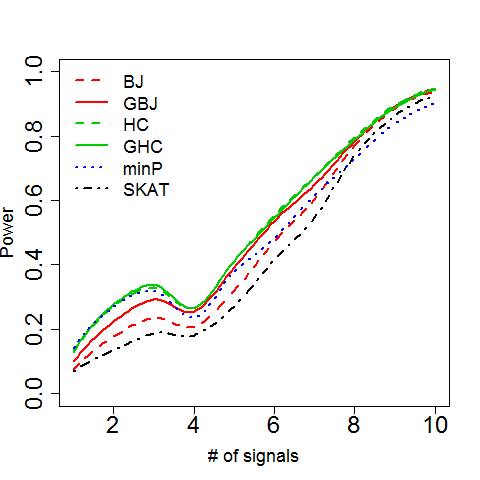}}
		\subfloat{\includegraphics[width=2.1in,height=2.2in]{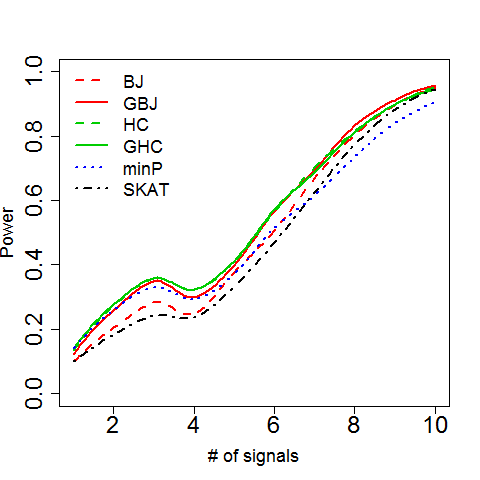}}
	\end{center}
	\caption{Power comparison under regression model with $N(0, 1)$ distributed covariates and polynomial-decay correlations. The left, middle, and right panels follow correlation structure Cases 5--7 in Table \ref{cor_types}, respectively, with $\gamma=1$. 
	}
\label{fig:power_linreg_poly}
\end{figure}


\begin{table}[H]
\centering
\caption{The effect sizes for different correlation structures and different number of signals.}
\label{effect_sizes}
\begin{tabular}{@{}ccccccccccc@{}}
\toprule
\# signals & 1     & 2     & 3     & 4     & 5     & 6     & 7     & 8     & 9     & 10    \\ \midrule
A          & 0.110 & 0.080 & 0.060 & 0.050 & 0.040 & 0.040 & 0.035 & 0.035 & 0.035 & 0.035 \\
B          & 0.110 & 0.080 & 0.060 & 0.050 & 0.040 & 0.040 & 0.035 & 0.035 & 0.035 & 0.035 \\
C          & 0.100 & 0.073 & 0.055 & 0.045 & 0.036 & 0.036 & 0.032 & 0.032 & 0.032 & 0.032 \\
D          & 0.085 & 0.062 & 0.046 & 0.038 & 0.031 & 0.031 & 0.027 & 0.027 & 0.027 & 0.027 \\
Case 1     & 0.120 & 0.100 & 0.090 & 0.090 & 0.090 & 0.090 & 0.090 & 0.090 & 0.090 & 0.090 \\
Case 2      & 0.110 & 0.080 & 0.060 & 0.050 & 0.040 & 0.040 & 0.035 & 0.030 & 0.030 & 0.030 \\
Case 3      & 0.110 & 0.090 & 0.060 & 0.050 & 0.050 & 0.040 & 0.030 & 0.030 & 0.030 & 0.030 \\
Case 4      & 0.100 & 0.070 & 0.050 & 0.040 & 0.035 & 0.030 & 0.025 & 0.025 & 0.025 & 0.025 \\
Case 5      & 0.110 & 0.080 & 0.060 & 0.050 & 0.040 & 0.040 & 0.035 & 0.030 & 0.030 & 0.030 \\
Case 6      & 0.110 & 0.080 & 0.060 & 0.050 & 0.040 & 0.040 & 0.035 & 0.030 & 0.030 & 0.030 \\
Case 7      & 0.110 & 0.080 & 0.060 & 0.050 & 0.040 & 0.040 & 0.035 & 0.030 & 0.030 & 0.030 \\ \bottomrule
\end{tabular}
\end{table}

\newpage
\vskip 0.2in
\bibliography{allMyReferences}

\begin{thebibliography}{32}
\providecommand{\natexlab}[1]{#1}
\providecommand{\url}[1]{\texttt{#1}}
\expandafter\ifx\csname urlstyle\endcsname\relax
  \providecommand{\doi}[1]{doi: #1}\else
  \providecommand{\doi}{doi: \begingroup \urlstyle{rm}\Url}\fi

\bibitem[Andr{\'e}s-Benito et~al.(2017)Andr{\'e}s-Benito, Moreno, Aso,
  Povedano, and Ferrer]{andres2017amyotrophic}
Pol Andr{\'e}s-Benito, Jes{\'u}s Moreno, Ester Aso, M{\'o}nica Povedano, and
  Isidro Ferrer.
\newblock Amyotrophic lateral sclerosis, gene deregulation in the anterior horn
  of the spinal cord and frontal cortex area 8: implications in frontotemporal
  lobar degeneration.
\newblock \emph{Aging}, 9\penalty0 (3):\penalty0 823--851, 2017.

\bibitem[Arias-Castro et~al.(2011)Arias-Castro, Cand{\`e}s, and
  Plan]{Arias2010}
E.~Arias-Castro, E.~J. Cand{\`e}s, and Y.~Plan.
\newblock Global testing under sparse alternatives: Anova, multiple comparisons
  and the higher criticism.
\newblock \emph{The Annals of Statistics}, 39\penalty0 (5):\penalty0
  2533--2556, 2011.

\bibitem[Barnett et~al.(2017)Barnett, Mukherjee, and
  Lin]{barnett2016generalized}
Ian Barnett, Rajarshi Mukherjee, and Xihong Lin.
\newblock The generalized higher criticism for testing snp-set effects in
  genetic association studies.
\newblock \emph{Journal of the American Statistical Association}, 112\penalty0
  (517):\penalty0 64--76, 2017.

\bibitem[Berk and Jones(1979)]{berk1979g}
R.~H. Berk and D.~H. Jones.
\newblock Goodness-of-fit test statistics that dominate the kolmogorov
  statistics.
\newblock \emph{Probability Theory and Related Fields}, 47\penalty0
  (1):\penalty0 47--59, 1979.

\bibitem[Bickel and Levina(2008)]{bickel2008regularized}
P.~J. Bickel and E.~Levina.
\newblock Regularized estimation of large covariance matrices.
\newblock \emph{The Annals of Statistics}, 36\penalty0 (1):\penalty0 199--227,
  2008.

\bibitem[Cai et~al.(2010)Cai, Zhang, and Zhou]{cai2010optimal}
T.~T. Cai, C.~H. Zhang, and H.~H. Zhou.
\newblock Optimal rates of convergence for covariance matrix estimation.
\newblock \emph{The Annals of Statistics}, 38\penalty0 (4):\penalty0
  2118--2144, 2010.

\bibitem[Cleveland et~al.(1992)Cleveland, Grosse, and Shyu]{cleveland1992local}
WS~Cleveland, E~Grosse, and WM~Shyu.
\newblock Local regression models. in `statistical models in s'.(eds jm
  chambers, tj hastie) pp. 309--376, 1992.

\bibitem[Donoho and Jin(2004)]{Donoho2004}
David~L Donoho and Jiashun Jin.
\newblock Higher criticism for detecting sparse heterogeneous mixtures.
\newblock \emph{The Annals of Statistics}, 32\penalty0 (3):\penalty0 962--994,
  2004.

\bibitem[Donoho and Jin(2008)]{Donoho2008}
David~L Donoho and Jiashun Jin.
\newblock Higher criticism thresholding: Optimal feature selection when useful
  features are rare and weak.
\newblock \emph{Proceedings of the National Academy of Sciences of the United
  States of America}, 105\penalty0 (39):\penalty0 14790--14795, Sep 30 2008.

\bibitem[Duerr et~al.(2006)Duerr, Taylor, Brant, Rioux, Silverberg, Daly,
  Steinhart, Abraham, Regueiro, Griffiths, et~al.]{Duerr2006}
R.H. Duerr, K.D. Taylor, S.R. Brant, J.D. Rioux, M.S. Silverberg, M.J. Daly,
  A.H. Steinhart, C.~Abraham, M.~Regueiro, A.~Griffiths, et~al.
\newblock A genome--wide association study identifies il23r as an inflammatory
  bowel disease gene.
\newblock \emph{Science Signalling}, 314\penalty0 (5804):\penalty0 1461, 2006.

\bibitem[Fan et~al.(2013)Fan, Jin, Yao, et~al.]{fan2013optimal}
Yingying Fan, Jiashun Jin, Zhigang Yao, et~al.
\newblock Optimal classification in sparse gaussian graphic model.
\newblock \emph{The Annals of Statistics}, 41\penalty0 (5):\penalty0
  2537--2571, 2013.

\bibitem[Hall and Jin(2010)]{Hall2010}
P.~Hall and J.~Jin.
\newblock Innovated higher criticism for detecting sparse signals in correlated
  noise.
\newblock \emph{The Annals of Statistics}, 38\penalty0 (3):\penalty0
  1686--1732, 2010.

\bibitem[Hugot et~al.(2001)Hugot, Chamaillard, Zouali, Lesage, C{\'e}zard,
  Belaiche, Almer, Tysk, O'Morain, Gassull, et~al.]{hugot2001association}
Jean-Pierre Hugot, Mathias Chamaillard, Habib Zouali, Suzanne Lesage,
  Jean-Pierre C{\'e}zard, Jacques Belaiche, Sven Almer, Curt Tysk, Colm~A
  O'Morain, Miquel Gassull, et~al.
\newblock Association of nod2 leucine-rich repeat variants with susceptibility
  to crohn's disease.
\newblock \emph{Nature}, 411\penalty0 (6837):\penalty0 599--603, 2001.

\bibitem[Jager and Wellner(2007)]{jager2007goodness}
Leah Jager and Jon~A Wellner.
\newblock Goodness-of-fit tests via phi-divergences.
\newblock \emph{The Annals of Statistics}, pages 2018--2053, 2007.

\bibitem[Jin and Ke(2016)]{jin2014rare}
Jiashun Jin and Zheng~Tracy Ke.
\newblock Rare and weak effects in large-scale inference: methods and phase
  diagrams.
\newblock \emph{Statistica Sinica}, 26:\penalty0 1--34, 2016.

\bibitem[Keogh and Chinnery(2013)]{keogh2013next}
MJ~Keogh and PF~Chinnery.
\newblock Next generation sequencing for neurological diseases: new hope or new
  hype?
\newblock \emph{Clinical neurology and neurosurgery}, 115\penalty0
  (7):\penalty0 948--953, 2013.

\bibitem[Kolmogorov(1933)]{komogorov1933}
Andrey Kolmogorov.
\newblock Sulla determinazione empirica di una leggi di distribuzione.
\newblock \emph{Giornale dell'Istituto Italiano degli Attuari}, 4:\penalty0
  83--91, 1933.

\bibitem[Kotz and Johnson(2012)]{kotz2012breakthroughs}
Samuel Kotz and Norman~L Johnson.
\newblock \emph{Breakthroughs in Statistics: Foundations and basic theory}.
\newblock Springer Science \& Business Media, 2012.

\bibitem[Li and Siegmund(2015)]{Li2014higher}
Jian Li and David Siegmund.
\newblock Higher criticism: $p$-values and criticism.
\newblock \emph{The Annals of Statistics}, 43\penalty0 (3):\penalty0
  1323--1350, 2015.

\bibitem[Littell and Folks(1971)]{littell1971asymptotic}
Ramon~C Littell and J~Leroy Folks.
\newblock Asymptotic optimality of {F}isher's method of combining independent
  tests.
\newblock \emph{Journal of the American Statistical Association}, 66\penalty0
  (336):\penalty0 802--806, 1971.

\bibitem[Price et~al.(2006)Price, Patterson, Plenge, Weinblatt, Shadick, and
  Reich]{Price2006}
A.~L. Price, N.~J. Patterson, R.~M. Plenge, M.~E. Weinblatt, N.~A. Shadick, and
  D.~Reich.
\newblock Principal components analysis corrects for stratification in
  genome-wide association studies.
\newblock \emph{Nature Genetics}, 38\penalty0 (8):\penalty0 904--909, 2006.

\bibitem[Shao(2010)]{shaostat}
J.~Shao.
\newblock \emph{Mathematical Statistics}.
\newblock Springer Verlag, 2010.

\bibitem[Shlyakhter et~al.(2014)Shlyakhter, Sabeti, and
  Schaffner]{shlyakhter2014cosi2}
Ilya Shlyakhter, Pardis~C Sabeti, and Stephen~F Schaffner.
\newblock Cosi2: an efficient simulator of exact and approximate coalescent
  with selection.
\newblock \emph{Bioinformatics}, 30\penalty0 (23):\penalty0 3427--3429, 2014.

\bibitem[Smith et~al.(2014)Smith, Ticozzi, Fallini, Gkazi, Topp, Kenna,
  Scotter, Kost, Keagle, Miller, et~al.]{smith2014exome}
Bradley~N Smith, Nicola Ticozzi, Claudia Fallini, Athina~Soragia Gkazi, Simon
  Topp, Kevin~P Kenna, Emma~L Scotter, Jason Kost, Pamela Keagle, Jack~W
  Miller, et~al.
\newblock Exome-wide rare variant analysis identifies {TUBA4A} mutations
  associated with familial {ALS}.
\newblock \emph{Neuron}, 84\penalty0 (2):\penalty0 324--331, 2014.

\bibitem[Stouffer et~al.(1949)Stouffer, Suchman, DeVinney, Star, and
  Williams]{Stouffer1949}
Samuel~A Stouffer, Edward~A Suchman, Leland~C DeVinney, Shirley~A Star, and
  Robin~M Williams.
\newblock \emph{The American Soldier: Adjustment during Army Life}, volume~I.
\newblock Princeton University Press, New Jersey, 1949.

\bibitem[Sun(2005)]{sun2005wiener}
Qiyu Sun.
\newblock Wiener's lemma for infinite matrices with polynomial off-diagonal
  decay.
\newblock \emph{Comptes Rendus Mathematique}, 340\penalty0 (8):\penalty0
  567--570, 2005.

\bibitem[Sun and Lin(2017)]{sun2017set}
Ryan Sun and Xihong Lin.
\newblock Set-based tests for genetic association using the generalized
  berk-jones statistic.
\newblock \emph{arXiv preprint arXiv:1710.02469}, 2017.

\bibitem[Tong(2012)]{tong2012multivariate}
Yung~Liang Tong.
\newblock \emph{The multivariate normal distribution}.
\newblock Springer Science \& Business Media, New York, 2012.

\bibitem[Wall and Pritchard(2003)]{wall2003haplotype}
Jeffrey~D Wall and Jonathan~K Pritchard.
\newblock Haplotype blocks and linkage disequilibrium in the human genome.
\newblock \emph{Nature Reviews Genetics}, 4\penalty0 (8):\penalty0 587, 2003.

\bibitem[Wu et~al.(2011)Wu, Lee, Cai, Li, Boehnke, and Lin]{wu2011rarevariant}
M.~C. Wu, S.~Lee, T.~Cai, Y.~Li, M.~Boehnke, and X.~Lin.
\newblock Rare-variant association testing for sequencing data with the
  sequence kernel association test.
\newblock \emph{American Journal of Human Genetics}, 89\penalty0 (1):\penalty0
  82--93, Jul 15 2011.

\bibitem[Wu et~al.(2014)Wu, Sun, He, Cho, Zhao, and Jin]{Wu2014detection}
Zheyang Wu, Yiming Sun, Shiquan He, Judy Cho, Hongyu Zhao, and Jiashun Jin.
\newblock Detection boundary and {H}igher {C}riticism approach for sparse and
  weak genetic effects.
\newblock \emph{The Annals of Applied Statistics}, 8\penalty0 (2):\penalty0
  824--851, 2014.

\bibitem[Zhang et~al.(2017)Zhang, Jin, and Wu]{Zhang2016distributions}
Hong Zhang, Jiashun Jin, and Zheyang Wu.
\newblock Distributions and statistical power of optimal signal-detection
  methods in finite cases.
\newblock \emph{arXiv:1801.04309.}, 2017.

\end{thebibliography}

\end{document}